\documentclass[webpdf,modern,medium,namedate]{oup-authoring-template}
\usepackage{booktabs}
\usepackage{longtable}
\providecommand{\XeTeXLinkBox}[1]{#1}

\onecolumn

\begin{document}

\pagestyle{plain}

\title[Emulation strategies for Bayesian inference of regional left ventricle material parameters]{Emulation strategies for Bayesian inference of regional left ventricle material parameters}

\author[1,$\ast$]{Hongjin Ren\ORCID{0009-0000-4242-0546}}
\author[1]{Vinny Davies\ORCID{0000-0003-1896-8936}}
\author[1]{Hao Gao\ORCID{0000-0001-6852-9435}}
\author[1]{Mu Niu\ORCID{0000-0002-3068-5501}}
\author[1]{Benn Macdonald\ORCID{0000-0003-3728-5305}}

\address[1]{\orgdiv{School of Mathematics and Statistics}, \orgname{University of Glasgow}, \orgaddress{\street{University Place}, \postcode{G12 8QQ}, \state{Glasgow}, \country{UK}}}

\corresp[$\ast$]{Corresponding author. \href{mailto:hongjin.ren@glasgow.ac.uk}{hongjin.ren@glasgow.ac.uk}}

\abstract{Patient-specific biomechanical models of the left ventricle can relate cardiac magnetic resonance imaging to regional myocardial material properties, but existing emulator-based studies typically treat the myocardium as mechanically homogeneous, limiting representation of localised dysfunction. We propose a Bayesian surrogate-modelling framework for inferring regional Holzapfel-Ogden material parameters in a left ventricle partitioned into five physiological zones derived from the American Heart Association 17-segment model. Eight emulator strategies spanning single- versus multi-output, local versus global, and Gaussian-process- versus neural-network-based architectures were screened using parameter point-estimation accuracy; the three retained models were evaluated using empirical marginal credible-interval coverage and posterior contraction. We found that models with comparable point accuracy nevertheless differed markedly in uncertainty. A multi-output variational Gaussian process provided the most favourable balance across these criteria and was retained for the subsequent analyses. In synthetic local and global stiffening scenarios, maximum a posteriori estimates generally distinguished stiffened from baseline zones, but the nonlinear-stiffening parameters were more difficult to identify from end-diastolic observations than the stiffness-magnitude parameters. A healthy-volunteer analysis demonstrates feasibility with an incomplete observation vector and jointly inferred strain-noise scales. These results suggest that the proposed framework provides a computationally feasible, uncertainty-aware approach to regional left ventricle parameter inference.}


\maketitle
\thispagestyle{plain}

\section{Introduction}
\label{sec:intro}

Quantifying how cardiac function differs between healthy and diseased hearts has long been a central objective in cardiovascular research. Patient specific biomechanical modelling, increasingly informed by cardiac magnetic resonance imaging (MRI), provides a framework for relating observed ventricular motion to underlying mechanical properties governing cardiac function. Recent studies, for example, have reported that peak muscle fibre stress in patients with obstructive hypertrophic cardiomyopathy (HCM) is lower than in healthy subjects \citep{nikouComputationalModelingHealthy2016}, whilst myocardial contractility in acute myocardial infarction (MI) patients may significantly exceed that of normal hearts \citep{gaoChangesClassificationMyocardial2017}. These findings highlight the importance of estimating subject specific mechanical behaviour from imaging data, particularly in the context of left ventricle (LV) systolic dysfunction, which remains a key therapeutic target \citep{clelandEpidemiologyManagementHeart2005}. Biomechanical models that infer such properties from ventricular motion provide a route to identifying how mechanical function varies across the heart and how these variations change under different pathological conditions \citep{emigPassiveMyocardialMechanical2021}.

Despite these clinical motivations, the use of biomechanical models in routine clinical settings remains limited. LV tissue mechanics are commonly described by constitutive laws, such as the widely used Holzapfel-Ogden (H-O) model, which incorporates myocardial fibre architecture to represent the strongly nonlinear and anisotropic mechanical behaviour of the myocardium through biologically related material parameters \citep{holzapfelConstitutiveModellingPassive2009}. A key difficulty is that these parameters are not directly observable in vivo and must instead be inferred by calibrating the model to patient data, typically by comparing model predicted quantities such as regional strains and cavity volume with measurements extracted from cardiac MRI and iteratively adjusting parameters to improve agreement. This calibration task forms a computationally intensive inverse problem in which repeated numerical solutions of the governing partial differential equations are required. Parameter estimation therefore typically involves many forward simulations, making direct simulation based approaches difficult to apply within clinically relevant time frames and limiting their suitability for routine use \citep{gaoParameterEstimationHolzapfel2015a}.

Statistical surrogate models, or emulators, offer an effective strategy to alleviate this computational bottleneck. An emulator is a regression based model trained on a limited set of precomputed simulations that learns to predict the output of a simulator across the parameter space at a fraction of the original cost. Importantly, emulator training can be performed offline i.e. prior to clinical use, with parameter spaces explored through space filling experimental designs such as Sobol sequences \citep{sobolDistributionPointsCube1967}. Gaussian process (GP) models have emerged as a particularly effective class of emulators for cardiac mechanics, as they provide flexible representations of simulator response surfaces while quantifying predictive uncertainty. Previous work has shown that GP based emulators can recover patient specific parameters with negligible loss of accuracy while substantially reducing computational cost relative to traditional numerical solution approaches \citep{daviesFastParameterInference2019a,noeGaussianProcessEmulation2019}. GP emulators have also been shown to outperform alternative machine learning surrogates, such as neural networks and random forests, in the data-sparse regimes typical of cardiac simulations \citep{daltonComparativeEvaluationDifferent2020a}. Consistent results have also been reported by \citet{renPerformanceComparisonStatistical2026} on simulated benchmark problems outside the biomechanical setting, where multi-output GP extensions exploiting correlations between outputs further improved emulation accuracy for spatially structured quantities such as distributed strain fields.

A critical limitation of most existing emulator based studies is that they treat the LV as a homogeneous structure, yielding a single global set of material parameters \citep{noeGaussianProcessEmulation2019}. Whilst global estimates are appropriate for characterising diffuse myocardial disease, they do not provide enough information to handle localised dysfunction, such as regional scarring following myocardial infarction, that is confined to specific portions of the ventricular wall \citep{mojsejenkoEstimatingPassiveMechanical2015,gaoEstimatingPrognosisPatients2017}. Modelling regional heterogeneity instead requires the estimation of separate parameter sets across multiple anatomical segments. In the present study, the LV wall is partitioned according to the 17 segment American Heart Association (AHA) scheme \citep{americanheartassociationwritinggrouponmyocardialsegmentationandregistrationforcardiacimaging:StandardizedMyocardialSegmentation2002} and subsequently grouped into five physiologically coherent regions: anterior, inferior, lateral, septal, and apical \citep{tatImpactLateGadolinium2022}. This regional parameterisation enlarges the parameter space relative to global estimation and introduces the possibility of practical non-identifiability, because distinct regional parameter combinations may produce similar end-diastolic strains and volume.

Previous studies have established the computational value of emulators for cardiac-mechanics parameter estimation, but less attention has been given to whether an emulator will also induce reliable posterior inference for regional parameters. Existing applications have mainly emphasised point estimation \citep{daviesFastParameterInference2019a} or posterior inference under a global parameterisation \citep{borowskaBayesianOptimisationEfficient2022,geAdvancedStatisticalInference2025}, which is not equivalent to the new challenge. In a Bayesian inverse problem, emulators with similar forward-prediction or point-estimation accuracy can induce different likelihoods and, consequently, different posterior coverage and contraction. Emulator selection for Bayesian inference should therefore consider the posterior distributions induced by the competing emulators rather than predictive accuracy alone.

The principal contribution is an inferentially targeted emulator assessment in which candidate models are compared using both predictive criteria and the properties of the parameter posteriors that they induce. Posterior dependence is interpreted as an exploratory description of compensatory parameter directions rather than as a formal characterisation of biomechanical identifiability. The present study focuses on a single LV geometry, providing a foundation for future extensions that incorporate patient-specific geometric variation, for example through principal component analysis (PCA)-based shape parameterisation as explored by \citet{lazarusImprovingCardioMechanicInference2022}. Such extensions would again increase the dimensionality of the inference problem and motivate further advances in emulation and dimension-reduction strategies.

The proposed Bayesian inference framework is evaluated in three stages. First, eight candidate emulators spanning single- versus multi-output, local versus global, and Gaussian-process- versus neural-network-based architectures are screened using parameter point-estimation accuracy. Second, the shortlisted emulators are embedded within an MCMC framework and assessed using convergence diagnostics, empirical marginal coverage, the continuous ranked probability score and posterior contraction. Third, the selected emulator is evaluated in four controlled scenarios representing local and global myocardial stiffening, followed by an application to cardiac MRI observations from a healthy volunteer. These analyses examine regional parameter recovery, posterior uncertainty and exploratory posterior dependence under synthetic and clinical observation settings.

Section~\ref{sec:LV} defines the biomechanical model and its outputs, while Sections~\ref{sec:Methodology} and~\ref{sec:data-experimental-design} describe the statistical framework and experimental designs. Section~\ref{sec:Results} reports the emulator comparison and the synthetic and clinical evaluations, followed by discussion and potential extensions in Sections~\ref{sec:Discussion} and~\ref{sec:Future work}.

\section{Biomechanical model of the left ventricle}
\label{sec:LV}

This section defines the forward biomechanical model of passive LV filling. A three-dimensional ventricular geometry reconstructed from cardiac MRI is represented as a fibre-reinforced hyperelastic material governed by the H-O constitutive law \citep{gaoQuasistaticImagebasedImmersed2014,holzapfelConstitutiveModellingPassive2009}. The myocardium is partitioned into five AHA-based physiological zones \citep{tatImpactLateGadolinium2022}, and passive filling is formulated as a quasi-static pressure-loaded boundary-value problem following \citet{gaoParameterEstimationHolzapfel2015a}. The resulting observational outputs are defined in Section~\ref{sec:QoI}.

\subsection{Patient-specific left ventricle geometry and regional partition}
\label{sec:left ventricle geometry}

The LV geometry used in this study is reconstructed from a cardiac MRI scan of a healthy volunteer obtained in our previous work and discretised using a three-dimensional finite element mesh \citep{gaoChangesClassificationMyocardial2017}. Images acquired at early diastole when the mitral valve just opens define the initial, approximately unloaded configuration of the ventricle, which is used as the reference configuration for the passive mechanical model.

\begin{figure}[!t]
\centering
\includegraphics[width=0.9\textwidth, keepaspectratio]{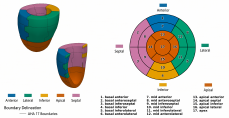}
\caption{Left ventricle geometry and five-zone parameterisation used throughout the study. The geometry was reconstructed from cardiac MRI of the study volunteer and discretised as a three-dimensional finite-element mesh in the early-diastolic reference configuration. Left panel shows full and cutaway mesh views coloured by the five zones used for parameter inference: anterior (blue), lateral (green), inferior (yellow), septal (pink), and apical (orange). Light lines indicate the underlying 17 AHA segments. Right panel shows the corresponding AHA bullseye map, with concentric rings denoting the basal, mid-cavity, apical, and apex levels.}
\figalttext{Two three-dimensional renderings of a left-ventricular mesh, base uppermost, one intact and one cut away to expose the cavity, sit beside a four-ring bullseye diagram. Four of the five zones form contiguous circumferential wedges running from base to mid-cavity; the fifth forms a continuous cap over the apical third, with septal opposite lateral and anterior opposite inferior. The bullseye rings carry the numbered segments, with the apex at the centre.}
\label{fig:lv-zones}
\end{figure}

The LV is partitioned according to the AHA 17-segment standard. Along the long axis, the basal and mid-cavity levels are each divided into six circumferential segments, the apical level into four segments, and the ventricular apex forms the apical cap. These 17 segments are grouped by anatomical location and functional similarity into five physiological zones associated with typical coronary perfusion territories: anterior, inferior, lateral, septal and apical \citep{tatImpactLateGadolinium2022}. The complete mapping is shown in Figure~\ref{fig:lv-zones}. All segments within a zone share the same pair of regional material parameters defined in Section~\ref{sec:H-O}. The five-zone representation retains regional anatomical structure while limiting the dimension that would result from assigning an independent parameter pair to every AHA segment.

\subsection{Myocardial constitutive law and passive filling}
\label{sec:H-O}

The passive mechanical response of the myocardium is described using the H-O constitutive law (\citealt{holzapfelConstitutiveModellingPassive2009}). The myocardium is treated as an incompressible, anisotropic and hyperelastic material whose strain-energy density consists of an isotropic matrix contribution, fibre and sheet reinforcement terms, and a fibre-sheet shear interaction term. The strain-energy density is written as
\begin{equation}
\Psi
=
\frac{a}{2b}\left[\exp\bigl(b(I_1-3)\bigr)-1\right]
+
\sum_{i\in\{f,s\}}
\frac{a_i}{2b_i}
\left[
\exp\bigl(b_i (\max(I_{4i},1)-1)^2\bigr)-1
\right]
+
\frac{a_{\mathrm{fs}}}{2b_{\mathrm{fs}}}
\left[
\exp\bigl(b_{\mathrm{fs}} I_{8\mathrm{fs}}^{2}\bigr)-1
\right].
\label{eq:holzapfel-ogden}
\end{equation}
Here $(a,b,a_{\mathrm{f}},b_{\mathrm{f}},a_{\mathrm{s}},b_{\mathrm{s}},a_{\mathrm{fs}},b_{\mathrm{fs}})$ are the eight material parameters of the full H-O constitutive law. The parameters $a$ and $b$ govern the isotropic matrix response, $a_{\mathrm{f}}$ and $b_{\mathrm{f}}$ describe reinforcement along the fibre direction, $a_{\mathrm{s}}$ and $b_{\mathrm{s}}$ describe reinforcement along the sheet direction, and $a_{\mathrm{fs}}$ and $b_{\mathrm{fs}}$ describe fibre-sheet shear coupling. The use of $\max(I_{4i},1)$ restricts the fibre and sheet contributions to extension rather than compression. The invariants are defined from the right Cauchy--Green deformation tensor $\mathbf{C} = \mathbf{F}^{\top}\mathbf{F}$, where \(\mathbf{F}\) is the deformation gradient. The relevant invariants are
\begin{equation}
\begin{aligned} 
I_1 &= \operatorname{tr}(\mathbf{C}), & \qquad I_{4\mathrm{f}} &= \mathbf{m}_0 \cdot \left(\mathbf{C}\mathbf{m}_0\right), \\ 
I_{4\mathrm{s}} &= \mathbf{s}_0 \cdot \left(\mathbf{C}\mathbf{s}_0\right), & \qquad I_{8\mathrm{fs}} &= \mathbf{m}_0 \cdot \left(\mathbf{C}\mathbf{s}_0\right), 
\end{aligned}
\end{equation}
where \(\mathbf{m}_0\) and \(\mathbf{s}_0\) denote the unit fibre and sheet directions in the reference configuration.

Studies have shown that cardiac MRI data is generally insufficient to support reliable inference of all eight parameters \citep{gaoParameterEstimationHolzapfel2015a}. Also, the sensitivity analyses report that end-diastolic cavity volume and circumferential strains are dominated by the matrix stiffness $a$ and the myofibres stiffness $a_{\mathrm{f}}$, the exponential parameters $b$ and $b_{\mathrm{f}}$ become identifiable mainly when high-pressure data are available, and the remaining parameters $(a_{\mathrm{s}}, b_{\mathrm{s}}, a_{\mathrm{fs}}, b_{\mathrm{fs}})$ have limited influence within the physiological pressure range \citep{lazarusSensitivityAnalysisInverse2022}. Therefore, several parameter-reduction strategies have been proposed in previous work to relieve this situation. \citet{gaoParameterEstimationHolzapfel2015a} introduced two scaling factors: one acting on an a-type group, representing stiffness magnitudes, and the other on a b-type group, representing exponential rates of nonlinear stiffening. This reduces the material parameter dimension from eight to two, with reference values for the underlying H-O parameters taken from ex vivo simple-shear fits to human and porcine myocardium \citep{wangStructureBasedFiniteStrain2013}. Both \citet{daviesFastParameterInference2019a} and \citet{noeGaussianProcessEmulation2019} adopted a four-factor reparametrization that pairs each $a$-parameter with its corresponding $b$-parameter according to its physical role.
 A different strategy was adopted by \citet{geAdvancedStatisticalInference2025}, who moved away from the scaling-factor formulation and fixed the less sensitive parameters $(a_{\mathrm{s}},b_{\mathrm{s}},a_{\mathrm{fs}},b_{\mathrm{fs}})$ at population values, while inferring only the four remaining parameters $(a,b,a_{\mathrm{f}},b_{\mathrm{f}})$. This preserves the original interpretation of the retained H-O parameters leading to a four-dimensional problem per anatomical region. 

In this paper, we adopt the two-factor reduction of \citet{gaoParameterEstimationHolzapfel2015a}, applied independently within each of the five AHA-based zones. This choice gives a tractable ten-dimensional regional parameter space while retaining the distinction between stiffness magnitude and nonlinear stiffening rate. The trade-off is that the relative ratios among the underlying H-O parameters are inherited from the reference parameter set rather than inferred separately for each patient. Specifically,
\begin{equation}
(a, a_{\mathrm{f}}, a_{\mathrm{s}}, a_{\mathrm{fs}})
=
C_a
(a_0, a_{\mathrm{f}0}, a_{\mathrm{s}0}, a_{\mathrm{fs}0}),
\qquad
(b, b_{\mathrm{f}}, b_{\mathrm{s}}, b_{\mathrm{fs}})
=
C_b
(b_0, b_{\mathrm{f}0}, b_{\mathrm{s}0}, b_{\mathrm{fs}0}),
\label{eq:parameter-scaling}
\end{equation}
where $C_a$ and $C_b$ are scaling factors and the subscript $0$ denotes population reference values. The factor $C_a$ scales the $a$-type stiffness magnitudes, while $C_b$ scales the $b$-type exponential stiffening, and $(a_0, a_{\mathrm{f}0}, a_{\mathrm{s}0}, a_{\mathrm{fs}0}, b_0, b_{\mathrm{f}0}, b_{\mathrm{s}0}, b_{\mathrm{fs}0})$ are reference values from the published literature of \citet{wangStructureBasedFiniteStrain2013}. Applying this reduction separately to the five zones gives the regional parameter vector
\begin{equation}
\boldsymbol{\theta}
=
\left(
C_a^{(1)}, C_b^{(1)},
C_a^{(2)}, C_b^{(2)},
\ldots,
C_a^{(5)}, C_b^{(5)}
\right)
\in \mathbb{R}^{10},
\label{eq:regional-parameter-vector}
\end{equation}
where the five pairs correspond to the anterior, inferior, lateral, septal and apical zones. This parameterisation preserves regional heterogeneity while reducing the dimension that must be explored during emulator-based Bayesian inference.

Passive LV filling is modelled as a quasi-static pressure-loaded boundary-value problem. The endocardial pressure is increased from zero to the prescribed end-diastolic pressure, and the resulting deformation field is obtained by solving the mechanical equilibrium equations. In the current LV domain $\Omega_t$, the problem can be written as
\begin{align}
\begin{cases}
\nabla \cdot \boldsymbol{\tau} = 0,
    & \text{in } \Omega_t, \\
\boldsymbol{\tau}\mathbf{n} = \mathbf{t},
    & \text{on } \Gamma^{\mathrm{endo}}, \\
u_r = u_z = 0,
    & \text{on } \Gamma^{\mathrm{base}},
\end{cases}
\label{eq:passive-filling-bvp}
\end{align}
where $\boldsymbol{\tau}$ is the Cauchy stress tensor, $\mathbf{n}$ is the outward unit normal on the endocardial surface $\Gamma^{\mathrm{endo}}$, $\mathbf{t}$ is the traction induced by the applied pressure, and $u_r$ and $u_z$ are displacement components on the basal surface $\Gamma^{\mathrm{base}}$.

The Cauchy stress is derived from the H--O strain-energy density as
\begin{equation}
\boldsymbol{\tau}
=
\mathbf{F}
\sum_{i=1,4\mathrm{f},4\mathrm{s},8\mathrm{fs}}
\frac{\partial \Psi}{\partial I_i}
\frac{\partial I_i}{\partial \mathbf{F}}
-
p \mathbf{I},
\label{eq:cauchy-stress-ho}
\end{equation}
where $p$ is the Lagrange multiplier enforcing incompressibility and $\mathbf{I}$ is the identity tensor. Solving this problem for a given $\boldsymbol{\theta}$ gives the end-diastolic (ED) LV configuration from which the output quantities are extracted.

\subsection{Quantities of interest}
\label{sec:QoI}

The forward model returns the end-diastolic LV cavity volume and segmental myocardial strains for each regional parameter vector $\boldsymbol{\theta}$. The end-diastolic LV cavity volume, denoted by $V_{\mathrm{ED}}$, is computed from the deformed endocardial surface at the end of passive filling. This quantity provides a global measure of LV chamber expansion under the prescribed end-diastolic pressure. 

Regional deformation is summarised by segmental circumferential, radial and longitudinal strains over the 17 AHA segments. Let $\mathbf{c}$, $\mathbf{r}$ and $\mathbf{l}$ denote the local unit circumferential, radial and longitudinal directions, respectively. The Green-Lagrange strain tensor used for post-processing is $\mathbf{E} = \frac{1}{2} \left( \widetilde{\mathbf{C}}-\mathbf{I} \right)$, where $\widetilde{\mathbf{C}} = \widetilde{\mathbf{F}}^{\top}\widetilde{\mathbf{F}}$ and $\widetilde{\mathbf{F}}$ denotes the deformation gradient with respect to the end-diastolic configuration, which is consistent with clinically measured strains. Equivalently, when $\mathbf{F}$ maps the early-diastolic reference configuration to the final end-diastolic configuration, one may write $\widetilde{\mathbf{F}} = \mathbf{F}^{-1}$.
The local strain components are then defined by 
$
\varepsilon_{\mathrm{cc}} = \mathbf{c}\cdot \left( \mathbf{E}\mathbf{c} \right), \varepsilon_{\mathrm{rr}} = \mathbf{r}\cdot \left( \mathbf{E}\mathbf{r} \right), \varepsilon_{\mathrm{ll}} = \mathbf{l}\cdot \left( \mathbf{E}\mathbf{l} \right). 
$
For each AHA segment $j=1,\ldots,17$, the reported segmental strain is the spatial average of the corresponding local strain component over that segment: 
$$ \bar{\varepsilon}_{q}^{(j)} = \frac{1}{|\Omega_j|} \int_{\Omega_j} \varepsilon_q(\mathbf{x})\,\mathrm{d}\mathbf{x}, \qquad q \in \{\mathrm{cc},\mathrm{rr},\mathrm{ll}\}, \quad j=1,\ldots,17, $$ 
where $\Omega_j$ denotes the myocardial region associated with AHA segment $j$. 

The full simulator output is therefore 
\begin{equation}
\mathbf{y}(\boldsymbol{\theta}) = \bigl( V_{\mathrm{ED}}, \bar{\boldsymbol{\varepsilon}}_{\mathrm{cc}}^{(j)}, \bar{\boldsymbol{\varepsilon}}_{\mathrm{rr}}^{(j)}, \bar{\boldsymbol{\varepsilon}}_{\mathrm{ll}}^{(j)} \bigr) \in \mathbb{R}^{52}, \quad j=1,\ldots,17,
\label{eq:simulator-output}
\end{equation}
These 52 quantities form the observational interface between the biomechanical simulator and the statistical inverse problem developed in the subsequent sections.

\section{Methodology}
\label{sec:Methodology}

This section presents the statistical framework used to infer the regional H-O constitutive parameters and quantify uncertainty in the resulting estimates. We begin in Section~\ref{sec:Simulation} by describing the underlying simulator and identifying the computational bottleneck that motivates surrogate modelling. Section~\ref{sec:Emulation} introduces the emulator concept, while Section~\ref{sec:training-output-representation} describes the training input design and the dimensionality reduction applied to the simulator output. Section~\ref{sec:emulator-strategies} then compares eight candidate emulator architectures organised along three architectural axes, before Section~\ref{sec:parameter-inference} presents the point estimation and Bayesian inference procedures used to evaluate each approach. Together, these components provide a statistically tractable surrogate framework for systematic inference using an otherwise computationally expensive biomechanical model.

\subsection{Simulation}
\label{sec:Simulation}

A simulator $\mathcal{M}$ is typically built on a complex mathematical model, such as the biomechanical models described in Section \ref{sec:LV}. The inputs of the simulator describe the physical properties of the system, and the outputs are observables that can be compared with measured data. In our setting, $\boldsymbol{\theta}\in\mathbb{R}^{10}$ encodes the two scaling parameters $(C_a^{(z)}, C_b^{(z)})$ for each of the five zones indexed by $z \in \{1,\ldots,5\}$, namely the Anterior, Inferior, Lateral, Septal, and Apical defined in Section~\ref{sec:H-O}. Given $\boldsymbol{\theta}$, the simulator returns the $52$-dimensional output vector $\mathbf{y}=\mathcal{M}(\boldsymbol{\theta})$ of Equation~\eqref{eq:simulator-output}. This forward simulator does not have a closed-form solution, and therefore a numerical solution is required to evaluate it each time the values of the parameters are changed. A single forward run requires approximately 20 minutes on our computing system (Intel Xeon Gold CPU 6138, 2.0-3.7GHz, 40 cores, and 128GB RAM). The inverse problem of estimating the unknown $\boldsymbol{\theta}$ that best reproduces the observed data $\mathbf{y}^{\mathrm{obs}}$, by minimising the discrepancy between $\mathbf{y}^{\mathrm{obs}}$ and $\mathcal{M}(\boldsymbol{\theta})$, typically requires thousands of forward evaluations under iterative optimisation or MCMC sampling. Direct use of the simulator within such an inferential framework is therefore computationally infeasible, motivating the surrogate-modelling strategy in the next section.

\subsection{Emulation}
\label{sec:Emulation}

An emulator $\hat{\mathcal{M}}$ is a statistical surrogate model trained on a finite set of pre-computed simulation datasets. It approximates the input--output mapping of $\mathcal{M}$ at substantially lower computational cost \citep{gramacySurrogatesGaussianProcess2020}. Once trained, an emulator prediction $\hat{\mathcal{M}}(\boldsymbol{\theta})$ can be obtained in milliseconds rather than minutes, allowing the inverse problem to become viable for an MCMC sampling framework. The emulator is therefore trained offline once, after which the resulting surrogate replaces iterative numerical solutions of $\mathcal{M}$ in downstream inferential tasks.

Constructing the emulator requires three related choices: the training locations used to cover the ten-dimensional parameter space $\Theta$, the representation of the strongly correlated simulator outputs, and the regression model used to approximate the resulting input--output relationship. The training design and output representation are common to all candidate emulators, whereas the regression architecture is the principal object of comparison.

\subsection{Training design and emulator output representation}
\label{sec:training-output-representation}

The emulator is trained on a finite set of input--output pairs $\{(\boldsymbol{\theta}_i,\mathbf{y}_i)\}_{i=1}^N$ and $\mathbf{y}_i=\mathcal{M}(\boldsymbol{\theta}_i)$. The high computational cost of each simulator evaluation motivates the use of a space-filling design for the input locations ${\boldsymbol{\theta}_i}$, with the aim of covering the parameter space $\Theta$ as evenly as possible under a fixed computational budget of $N$ evaluations. We use the Sobol sequence in this paper, a deterministic low-discrepancy sequence that provides approximate uniform coverage of $[0,1]^D$ for any prefix length $N$ and is widely used in computer experiments \citep{garudDesignComputerExperiments2017}. In this setting, the Sobol design is used to obtain a low-discrepancy set of input locations over $\Theta$, improving space-filling behaviour relative to purely random sampling under the same computational budget. In addition, to improve coverage in the low-stiffness region of $\Theta$, where the simulator output is most sensitive to small parameter perturbations, we augment the natural-space Sobol design with a second Sobol design generated in the log-transformed parameter space \citep{lazarusSensitivityAnalysisInverse2022}. The numerical details of the hybrid design, including the number of natural-space and log-space samples and the test-set size used for emulator validation, are reported in Section~\ref{sec:data-experimental-design}.

The simulator output $\mathbf{y}$ comprises 51 regional strain components and one end-diastolic volume scalar. Directly emulating all 52 outputs, ignoring the inter-output correlation arising from mechanical coupling within myocardial zone boundaries, would make the emulator training inefficient and could lead to numerical instability. We apply PCA to the standardised strain vector $\mathbf{s}(\boldsymbol{\theta})\in\mathbb{R}^{51}$ and retain the first $k$ principal components $\mathbf{z}(\boldsymbol{\theta})\in\mathbb{R}^{k}$. The end-diastolic volume scalar $V_{\mathrm{ED}}$ is retained without reduction, because it is one-dimensional and physically distinct from the strain components. The emulator is therefore trained on the reduced output 
$\tilde{\mathbf{y}}(\boldsymbol{\theta})=\bigl\{\mathbf{z}(\boldsymbol{\theta})^{\top},\;V_{\mathrm{ED}}\bigr\}^{\top}\in\mathbb{R}^{k+1}.$
PCA provides a computationally inexpensive, linear representation that captures the dominant directions of variation in the simulator outputs. After fitting PCA to the training outputs, the resulting mean vector and loading matrix are fixed and used to project both training and test outputs into the same reduced space. This reduces the emulation target from the full output vector to a small number of scores, allowing the emulator to focus on the potentially nonlinear mapping from the material parameters to those scores.

\subsection{Emulator strategies}
\label{sec:emulator-strategies}

The remaining design decision, and the main methodological focus of this paper, is the choice of regression model used to construct the emulator. One natural starting point is GP regression, a non-parametric Bayesian approach that returns, at each test point, a predictive mean that can be used as a fast surrogate for the simulator output, together with a predictive variance that quantifies emulator uncertainty \citep{rasmussenGaussianProcessesMachine2005}. This uncertainty is useful for inverse inference because it can be incorporated into the likelihood, allowing emulator uncertainty to contribute to posterior uncertainty. Standard GP training, however, scales as $\mathcal{O}(N^3)$ in the number of training simulations, and prediction can also become costly when the emulator is evaluated repeatedly inside an MCMC algorithm. The candidate emulators draw on three base regression models. The first is exact GP regression, as introduced above. The second is variational GP (VGP) regression, a sparse approximation in which the full GP is represented through inducing variables and variational inference. This retains an explicit estimate of predictive uncertainty while making global GP training more computationally feasible. The third is a feed-forward neural network (NN), a fully parametric regressor that scales readily to large training sets but, in the implementation considered here, returns only point predictions and therefore provides no intrinsic measure of emulator uncertainty.

\citet{daviesFastParameterInference2019a} demonstrated the viability of GP emulation for cardiac mechanics by treating the LV as mechanically homogeneous and inferring constitutive parameters in a four-dimensional parameter space, achieving an approximately three-orders-of-magnitude reduction in computational cost relative to direct simulation-based inference. The present study considers a regional ten-dimensional parameterisation, making three modelling questions more prominent. The first is whether computational cost is better controlled by restricting training locally or by using a scalable global approximation. The second is whether the emulator should represent cross-output dependence or emulate each output component independently. This is relevant because the volume and strain outputs are generated by the same mechanical deformation and may contain shared information about the underlying material parameters. The third is whether deep-learning components, such as learned feature extractors, hierarchical latent layers, or a fully neural-network emulator, improve predictive accuracy at acceptable computational cost. We address these questions by comparing eight candidate emulators that combine selected instances of these modelling choices, as summarised in Table~\ref{tab:emulator-architectures}.

\subsubsection{Training scope}
\label{sec:local-global}

The first question concerns whether the GP should be trained on the full training set of $N$ inputs or on a problem-adapted subset. Standard GP training requires an $\mathcal{O}(N^3)$ covariance factorisation, dominated by operations on the $N\times N$ training covariance matrix. With $N$ on the order of several thousand simulations, this cost is expensive for workflows that require repeated likelihood evaluations, such as MCMC sampling.

Two strategies are considered to address this drawback. The local strategy, adopted by \citet{daviesFastParameterInference2019a} and \citet{noeGaussianProcessEmulation2019}, restricts the training set to the $K$ nearest neighbours (KNN) of each test output in the output space. It therefore replaces the full $N\times N$ covariance matrix with a $K\times K$ matrix and reduces the local fitting cost to $\mathcal{O}(K^3)$. The advantage of the local approach is that the training subset adapts to each test point, allowing the emulator to respond to local features of the response surface. The disadvantage is that no readable global representation of the response surface is maintained, and predictive variance may be unstable in regions where the distance to the $K$-th nearest neighbour changes sharply across $\Theta$.

The global strategy, by contrast, retains the full training set and obtains scalability through a sparse variational approximation which is variational GP \citep{damianouVariationalGaussianProcess2011}. A set of $M$ inducing points, with $M\ll N$, is introduced as a low-rank summary of the training-set covariance, reducing the dominant training cost to $\mathcal{O}(NM^2)$ under common sparse variational formulations. Inducing-point locations and kernel hyperparameters are jointly optimised by maximising a variational lower bound on the log marginal likelihood. The advantage of the global approach is that it returns a single coherent posterior over the entire response surface, with smoothly varying predictive variance and a fixed offline training cost.

\subsubsection{Output structure}
\label{sec:single-multi-output}

The second architectural question concerns whether the reduced outputs should be modelled independently or jointly. The single-output strategy fits an independent GP to each component of the reduced output vector, ignoring cross-output dependence. Its advantage is computational simplicity, since the independent GPs can be fitted in parallel. Its disadvantage is that it cannot borrow information across related outputs which means when several outputs are constrained by the same mechanical deformation, this independence assumption may lead to less efficient predictions or less well-calibrated uncertainty.

The multi-output strategy instead models the outputs jointly. In a GP setting, this requires a covariance structure that describes both similarity between input locations and dependence between output components. A simple coregionalisation construction is the intrinsic coregionalisation model (ICM), under which the cross-covariance between outputs $y_k$ and $y_l$ at inputs $\boldsymbol{\theta}_i$ and $\boldsymbol{\theta}_j$ is
\begin{equation}
\operatorname{cov}(y_k(\boldsymbol{\theta}_i),y_l(\boldsymbol{\theta}_j))
=b_{kl}\,K(\boldsymbol{\theta}_i,\boldsymbol{\theta}_j),
\label{eq:icm-cov}
\end{equation}
where $K(\cdot,\cdot)$ is a covariance kernel over the input space shared across all outputs, and $\mathbf{B}=[b_{kl}]\in\mathbb{R}^{d\times d}$ is a symmetric positive-semidefinite coregionalisation matrix, where $d$ denotes the number of outputs \citep{bonillaMultitaskGaussianProcess2007}. The diagonal entries of $\mathbf{B}$ control the marginal variance of each output, while the off-diagonal entries encode pairwise correlation strength. The advantage of ICM is that information observed at one output can sharpen predictions at correlated outputs, however, it forces all outputs to share common smoothness and length-scale characteristics.

The linear model of coregionalisation (LMC) relaxes this constraint by introducing $Q$ base kernels:
\begin{equation}
\operatorname{cov}(y_k(\boldsymbol{\theta}_i),y_l(\boldsymbol{\theta}_j))
=\sum_{q=1}^{Q}b_{kl}^{(q)}\,K_q(\boldsymbol{\theta}_i,\boldsymbol{\theta}_j),
\label{eq:lmc-cov}
\end{equation}
where each $K_q(\cdot,\cdot)$ is a base kernel with its own hyperparameters, and $\mathbf{B}^{(q)}$ is the coregionalisation matrix associated with the $q$-th component \citep{alvarezComputationallyEfficientConvolved2011}. This allows different latent components to represent different input length-scales and different patterns of output dependence, at the cost of additional kernel hyperparameters and coregionalisation matrices.

The choice between ICM and LMC is linked to the training scope. The local multi-output emulators, L.MGP and L.DKMGP, fit a coregionalised GP on a nearest-neighbour subset, and this fit is repeated as the query input changes. For these local models, we use an ICM-style covariance structure, since the single shared kernel keeps each local fit computationally manageable. The multi-output variational GP, MVGP, is trained once on the full dataset and can therefore use the more flexible LMC representation. This allows the global multi-output variational model to represent more than one latent pattern of cross-output dependence, while still retaining a scalable variational formulation. The covariance kernel was selected from radial basis function (RBF), Matérn and rational quadratic (RQ) candidates using $K$-fold cross-validation within the simulator training set. All candidate kernels used automatic relevance determination (ARD), allowing a separate length-scale parameter for each input dimension.

\subsubsection{Deep component}
\label{sec:deep-extensions}

The third question concerns whether deep-learning components can improve predictive accuracy at acceptable computational cost. We consider three candidate extensions, each treated separately below.

The DNN emulator approximates the simulator output through a fully parametric feed-forward architecture. The advantage of the DNN is that it scales efficiently to large training sets and is well suited to capturing strongly nonlinear input-output mappings. The disadvantage is that, in its standard form, the DNN returns only a deterministic point prediction and does not provide intrinsic quantification of emulator uncertainty. To embed the DNN into the same Bayesian inference framework as the GP variants, we introduce a single error-scale parameter $\sigma$, sampled jointly with the constitutive parameters within the MCMC sampler of Section~\ref{sec:parameter-inference}. This parameter absorbs the combined effects of model approximation error and residual observation noise.

Deep kernel learning (DKL) combines a neural network and a GP by passing the input through a feature-extracting network and fitting a GP in the resulting feature space \citep{wilsonDeepKernelLearning2015}. The network parameters and kernel hyperparameters are jointly optimised by maximising the log marginal likelihood. The advantage of DKL relative to a standard GP is that the network can capture nonlinear input structure that a stationary kernel may miss, while the GP aspect retains uncertainty quantification. The disadvantage is the additional computational overhead introduced by joint network-kernel optimisation. For this reason, we combine DKL only with the local GP framework of Section~\ref{sec:local-global}, where the local subset keeps optimisation tractable.

A deep Gaussian process (DGP) replaces the deterministic feature extractor of DKL with a hierarchy of VGPs, so that intermediate representations are themselves probabilistic \citep{damianouDeepGaussianProcesses2013}. The advantage of the DGP is that it provides a probabilistic hierarchical representation rather than a deterministic feature map. The disadvantage is that training cost grows rapidly with depth, and intermediate covariance matrices can become numerically ill-conditioned in deeper configurations. We therefore adopt a two-layer DGP throughout, since deeper configurations were computationally impractical and numerically unstable in our setting.

\subsection{Parameter inference}
\label{sec:parameter-inference}

We now turn to inference of the regional H-O parameters $\boldsymbol{\theta}\in\mathbb{R}^{10}$ from observed data $\mathbf{y}^{\mathrm{obs}}$. We not only calculate point estimates $\hat{\boldsymbol{\theta}\hspace{0.05cm}}$ suitable for rapid reporting, but also derive a full posterior distribution $\pi(\boldsymbol{\theta}\mid\mathbf{y}^{\mathrm{obs}})$, which helps quantify the uncertainty arising from observational noise, emulator approximation error, and the correlation of the parameter space. The posterior is the principal inferential output of this paper, and point estimates are reported alongside it for reference.

\subsubsection{Point estimation}
\label{sec:point-estimation}

Point estimates are obtained by minimising the discrepancy between the emulator prediction $\hat{\mathcal{M}}(\boldsymbol{\theta})$ and the observation $\mathbf{y}^{\mathrm{obs}}$:
\begin{equation}
\hat{\boldsymbol{\theta}}
=\arg\min_{\boldsymbol{\theta}}\left\|\hat{\mathcal{M}}(\boldsymbol{\theta})-\mathbf{y}^{\mathrm{obs}}\right\|,
\label{eq:point-estimation}
\end{equation}
where $\hat{\mathcal{M}}(\boldsymbol{\theta})$ is the emulator prediction at $\boldsymbol{\theta}$, equal to the predictive mean $\hat{\boldsymbol{\mu}}(\boldsymbol{\theta})$ for the GP-based emulators and to the network output for the DNN, and $\mathbf{y}^{\mathrm{obs}}$ is the corresponding observed output vector. We use a two-stage optimisation procedure combining basinhopping and Adam. The initial parameter vectors are generated using a Sobol sequence providing space-filling starting values throughout the admissible parameter region. Basinhopping is used to reduce the risk of convergence to a local optimum by repeatedly perturbing the current solution and exploring different basins of the objective surface, and Adam is used for gradient-based optimisation of the emulator objective. This objective function uses the same emulator mapping and observation vector as the Bayesian formulation in the next section, but reduces the inverse problem to a least-squares criterion based only on the predictive mean. It therefore ignores both the predictive covariance and the prior distribution on $\boldsymbol{\theta}$. The advantage of this optimisation-based estimate is that it provides a computationally inexpensive deterministic summary of the inverse problem. The disadvantage is that it provides no measure of inferential uncertainty, which is addressed by the Bayesian formulation that follows.

\subsubsection{Bayesian inference}
\label{sec:bayesian-inference}

Point estimates alone are insufficient when distinct parameter configurations may produce nearly identical simulator outputs. To quantify this uncertainty, we use a Bayesian approach under a uniform prior $p(\boldsymbol{\theta})$ on the parameter bounds. The posterior distribution of $\boldsymbol{\theta}$ is given by Bayes' theorem:
\begin{equation}
p(\boldsymbol{\theta}\mid\mathbf{y}^{\mathrm{obs}})
\propto p(\mathbf{y}^{\mathrm{obs}}\mid\boldsymbol{\theta})\,p(\boldsymbol{\theta}).
\label{eq:bayes-theorem}
\end{equation}

The likelihood $p(\mathbf{y}^{\mathrm{obs}}\mid\boldsymbol{\theta})$ depends on the emulator architecture under consideration. 

In a single-output GP emulator, each output dimension is modelled independently and the predictive covariance is diagonal. The likelihood factorises as
\begin{equation}
p(\mathbf{y}^{\mathrm{obs}}\mid\boldsymbol{\theta})
=\prod_{k=1}^{d}\mathcal{N}(y_k^{\mathrm{obs}}|\hat{\mu}_k(\boldsymbol{\theta}),\hat{\sigma}_k^2(\boldsymbol{\theta})),
\label{eq:single-output-gp-likelihood}
\end{equation}
where $\hat{\mu}_k(\boldsymbol{\theta})$ and $\hat{\sigma}_k^{2}(\boldsymbol{\theta})$ are the predictive mean and variance returned by the $k$-th independent GP.

In a multi-output GP emulator, the predictive covariance is non-diagonal, and the likelihood retains the full multivariate Gaussian form:
\begin{equation}
p(\mathbf{y}^{\mathrm{obs}}\mid\boldsymbol{\theta})
=\mathcal{N}(\mathbf{y}^{\mathrm{obs}};\hat{\boldsymbol{\mu}}(\boldsymbol{\theta}),\hat{\boldsymbol{\Sigma}}(\boldsymbol{\theta})),
\label{eq:multi-output-gp-likelihood}
\end{equation}
where $\hat{\boldsymbol{\mu}}(\boldsymbol{\theta})=\{\hat{\mu}_1(\boldsymbol{\theta}),\ldots,\hat{\mu}_d(\boldsymbol{\theta})\}^{\top}\in\mathbb{R}^{d}$ is the predictive mean vector, and $\hat{\boldsymbol{\Sigma}}(\boldsymbol{\theta})\in\mathbb{R}^{d\times d}$ is the predictive covariance matrix returned by the multi-output GP. The diagonal of $\hat{\boldsymbol{\Sigma}}(\boldsymbol{\theta})$ contains the per-output predictive variances, while the off-diagonal entries encode the predicted correlations between outputs. Compared with the single-output likelihood, the multi-output likelihood retains inter-output correlation information and therefore imposes additional structure on $\boldsymbol{\theta}$.

For the DNN emulator considered here, predictive uncertainty is not modelled explicitly. A shared error-scale parameter $\sigma$ is therefore introduced with a half-normal prior, $\sigma\sim\mathrm{HalfNormal}(s)$. This parameter is sampled jointly with $\boldsymbol{\theta}$ within the MCMC framework, capturing the combined effects of emulator approximation error and residual observation noise. The resulting DNN likelihood is
\begin{equation}
p(\mathbf{y}^{\mathrm{obs}}\mid\boldsymbol{\theta},\sigma)
=\prod_{k=1}^{d}\mathcal{N}(y_k^{\mathrm{obs}};\hat{\mathcal{M}}_k(\boldsymbol{\theta}),\sigma^2),
\label{eq:dnn-likelihood}
\end{equation}
where $\sigma^{2}$ is a single scalar that does not vary with $\boldsymbol{\theta}$ or across output dimensions.

Posterior samples are drawn using the No-U-Turn Sampler (NUTS; \citealt{hoffmanNoUTurnSamplerAdaptively2014}), implemented within the Pyro probabilistic programming framework \citep{binghamPyroDeepUniversal2019}. NUTS extends Hamiltonian Monte Carlo by adaptively determining the trajectory length through a recursive doubling procedure, avoiding the need to specify this parameter manually while enabling efficient exploration of the posterior distribution.

Two independent chains are run for each test sample, each chain discards 300 warm-up iterations and retains 600 posterior draws, yielding a posterior sample set $\{\boldsymbol{\theta}^{(1)},\ldots,\boldsymbol{\theta}^{(M)}\}$ of size $M=1,200$. One exception was made for L.DKMGP, whose posteriors contracted least relative to the prior, so the number of draws per chain was increased from $600$ to $750$ to check that this behaviour reflected the likelihood induced by that emulator rather than an insufficient sampling budget. This is a modest number of retained draws, but was chosen as a compromise between Monte Carlo accuracy and the computational cost of repeated emulator-based likelihood evaluations. The same sampler settings are used for the simulated test cases and the clinical application, and convergence is assessed using the split-$\hat{R}$ statistic and the effective sample size (ESS). In the clinical-data application, additional noise-scale parameters representing the magnitude of measurement noise on the strain channels are sampled jointly within the same framework, allowing the data to inform the degree of observational uncertainty.

\section{Data and experimental design}
\label{sec:data-experimental-design}

This section describes the data used to train the emulators and to evaluate the inference framework presented in Section~\ref{sec:Methodology}. The training and emulator-validation simulations, generated using a hybrid Sobol design, are described in Section~\ref{sec:training-design}. Simulated disease scenarios used to assess inference performance under synthetic pathological conditions are presented in Section~\ref{sec:simulated-disease-scenarios}. Finally, a clinical dataset acquired from a healthy volunteer is introduced in Section~\ref{sec:clinical-data} to demonstrate the framework using cardiac magnetic resonance imaging measurements.

\subsection{Training and test design}
\label{sec:training-design}

Following the design in Section~\ref{sec:training-output-representation}, a Sobol sequence was generated over $\Theta=[0.1, 5]^{10}$, with parameter bounds informed by the parameter ranges reported by \citet{gaoParameterEstimationHolzapfel2015a}. Each configuration was evaluated using the simulator described in Section~\ref{sec:QoI}. Configurations for which the nonlinear solver failed to converge were excluded from the training set, since they did not provide valid simulator outputs. Such failures were typically associated with extreme or numerically challenging parameter combinations, for which the passive-filling problem became ill-conditioned or failed to reach a stable equilibrium within the solver tolerances. This screening retained $4,196$ converged simulations, providing broad coverage of $\Theta$ on its natural scale. A second, shorter Sobol sequence was then generated on the logarithmic scale and passed through the same convergence screening, contributing a further $1,465$ converged simulations. Since uniform sampling on the logarithmic scale places relatively more points at smaller values on the natural scale, this second batch increases the density of training data in the lower-parameter region, which is closer to healthy reference values \citep{lazarusSensitivityAnalysisInverse2022}. The two batches together provide $N=5,661$ training simulations, combining broad space-filling coverage of $\Theta$ on the natural scale with additional resolution in a clinically relevant lower-parameter regime.

A further $120$ test simulations were generated from an independent Sobol sequence over $\Theta$, held out from training, and used only for emulator selection and comparison. This test set is kept independent of the simulated disease scenarios introduced in Section~\ref{sec:simulated-disease-scenarios}, since the two test sets serve different methodological purposes. This dataset of $120$ simulations is used to select the emulator, whereas the disease scenario in Section~\ref{sec:simulated-disease-scenarios} evaluates the performance of the parameter inference.

\subsection{Simulated disease scenarios}
\label{sec:simulated-disease-scenarios}

We designed four simulated disease scenarios, structured around two parts: local and global stiffening, and the severity of material stiffness increase. This design enables us to assess, separately, the ability to localise the disease and its sensitivity to disease severity.

There are three scenarios that represent potential local myocardial disease due to stiffening, where increased stiffness is localised to a single anatomical territory. In each local scenario, only one of the five regions is affected. The constitutive scaling parameters in the four healthy regions are not fixed exactly at the population baseline value, but are sampled around it, with both $C_a$ and $C_b$ drawn from a normal distribution $\mathcal{N}(1,\,0.05)$. This provides an assessment of whether the inference framework can distinguish a genuinely stiffened region from ordinary inter-regional variability. The scaling parameters of the stiffened region are shifted to one of three increasing severity levels, $\mu \in \{1.5,\,2.0,\,3.0\}$, where $\mu$ denotes the mean of the sampling distribution for both $C_a$ and $C_b$ in the affected region. Thus, the corresponding scaling parameters are sampled from $\mathcal{N}(\mu,\,0.5)$ for the stiffened region, representing mild, moderate, and severe local stiffening. Each local scenario comprises 100 test samples, with 20 samples generated for each of the five potential diseased regions, ensuring that no single anatomical zone is over-represented.

The remaining scenario represents global myocardial stiffening, as this may occur in some forms of cardiomyopathy or diffuse fibrosis. In this scenario, the scaling parameters in all five regions are sampled from a normal distribution with $\mathcal{N}(1.7,\,0.5)$. This represents diffuse but not extreme stiffening of the myocardium across all anatomical territories. We refer to this as the global stiffening scenario and a total of $100$ test samples are generated under this scenario.

True parameter values are known by construction in all four scenarios, permitting direct evaluation of point estimation accuracy, posterior calibration, and credible-interval coverage, however they are not used to guide any inference procedure. The complete simulated test cases therefore comprise of $4 \times 100 = 400$ samples spanning a range of pathological representations.

\subsection{Clinical data}
\label{sec:clinical-data}

Clinical data were acquired from a cardiac MRI scan of a healthy volunteer \citep{gaoChangesClassificationMyocardial2017}. The available observation vector is comprised of the LV end-diastolic cavity volume together with 32 segmental strain measurements: 16 circumferential and 16 radial strains across the basal, mid-cavity, and apical short-axis slices, with the apical-cap segment omitted from the image-processing protocol.

The clinical application requires two modifications to the inference framework. The clinical observation vector is lower-dimensional than the simulator output. Whereas the simulator returns 52 quantities (see Section~\ref{sec:Simulation}), the clinical dataset contains only 33 outputs comprised of the cavity volume and 32 strain measurements. Longitudinal strains are unavailable, and the apical-cap segment is absent from the strain map. Consequently, the likelihood described in Section~\ref{sec:bayesian-inference} is evaluated using the corresponding 33-dimensional subset of the simulator output rather than the full 52-dimensional output. This reduction in observational information increases the inferential challenge by providing fewer constraints on the same ten-dimensional parameter vector.

In addition, the measurement-noise variance is unknown for the clinical data. To accommodate this, two additional parameters, $\sigma_1$ and $\sigma_2$, representing the standard deviations of the circumferential and radial measurement errors, respectively, are introduced into the inference model and sampled jointly with the constitutive parameter vector $\boldsymbol{\theta}$ using the No-U-Turn Sampler described in Section~\ref{sec:bayesian-inference}. The resulting posterior distribution is therefore \mbox{12-dimensional}, while the cavity-volume observation is treated as exact owing to its relatively high measurement reliability.

Jointly inferring $\sigma_1$ and $\sigma_2$ allows the observational uncertainty to be estimated from the data rather than imposed a priori, but at the cost of increasing the dimensionality of the inference problem and placing additional demands on parameter identifiability. This issue is revisited in Section~\ref{sec:Clinical data application}, where the clinical posterior summaries are reported alongside the inferred noise parameters and associated posterior diagnostics.

\section{Results}
\label{sec:Results}

This section reports the observed evaluation on the simulator outputs and the clinical observations. Section~\ref{sec:results-pca} shows the dimensionality reduction applied to the simulator output. Section~\ref{sec:results-emulator-selection} selects the primary emulator through a two-stage screening procedure, in which a Bayesian inference assessment of the shortlisted emulators follows an initial comparison based on parameter point estimation accuracy. Section~\ref{sec:results-stiffness} evaluates inference under simulated regional material stiffness, with the primary objective of distinguishing stiffened regions from baseline regions, and then characterises the dependence of inferential performance on stiffness severity, anatomical zone, and posterior correlation structure. Section~\ref{sec:Clinical data application} applies the framework to clinical cardiac magnetic resonance observations from a healthy volunteer.

\subsection{Output dimensionality reduction}
\label{sec:results-pca}

\begin{figure}[t]
  \centering
  \includegraphics[width=0.85\textwidth, keepaspectratio]{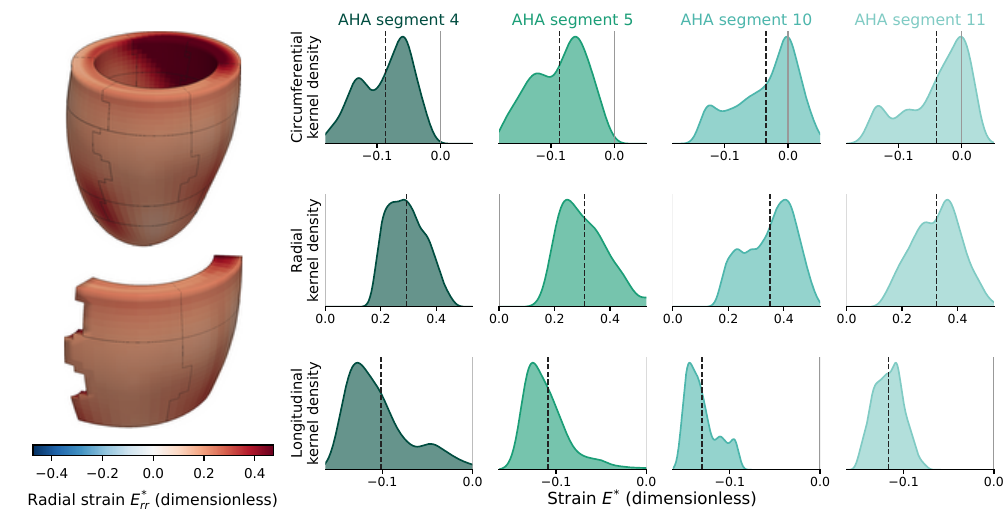}
  \caption{Visualisation of Radial strain in the left ventricle and lateral wall strain distributions. The top-left panel shows the Radial strain field over the full left ventricle geometry. The bottom-left panel shows the Radial strain in the lateral wall, defined here as AHA segments 4, 5, 10 and 11. Kernel density estimates of circumferential strain \(E_{cc}\), radial strain \(E_{rr}\), and longitudinal strain \(E_{\ell\ell}\) within the AHA segments comprising the lateral wall are shown on the right panel. Vertical dashed lines denote the segment-wise median.}
  \figalttext{Two shaded renderings of a left-ventricular mesh, one whole and one a cut-out portion of wall, sit beside a grid of filled density curves three rows deep and four columns wide. Every circumferential and longitudinal curve lies wholly below zero and every radial curve wholly above it. The radial curves are roughly three times as broad as the other two components, and each circumferential curve carries a secondary shoulder left of its main peak.}
  \label{fig:strain_within_group}
\end{figure}

The training simulations confirm the strain redundancy anticipated in Section~\ref{sec:training-output-representation}. Taking the lateral wall as an example, the strain distributions across its four constituent AHA segments are closely similar, as shown in Figure~\ref{fig:strain_within_group}. This pattern is consistent with shared regional deformation under passive filling and motivates a reduced representation of the strain outputs before emulator training.

If the segmental strains were modelled directly, this within-zone redundancy would lead to a high-dimensional output vector with strong linear dependence between components. This is a direct consequence of the regional output and is less prominent in mechanically homogeneous formulations, where the output representation is lower-dimensional. This near-collinearity can make covariance and coregionalisation estimates less stable when fitting the multi-output emulators. We therefore apply PCA to the standardised strain vector. The retained dimension was chosen as the smallest $k$ for which the cumulative proportion of explained strain variance exceeded $99.9\%$; retaining these components together with the volume scalar reduces the emulator output dimension from $52$ to $21$. The resulting representation preserves the dominant strain variation while reducing the inter-segment linear dependence passed to the emulator.

\subsection{Emulator selection}
\label{sec:results-emulator-selection}

We compared the eight candidate emulators introduced in Section~\ref{sec:emulator-strategies} using a two-stage selection. The first stage filters the eight candidates using parameter-space mean squared error (MSE) on $120$ held-out test samples, the second stage assesses uncertainty quantification under the No-U-Turn Sampler scheme of Section~\ref{sec:bayesian-inference} for the shortlisted emulators. The continuous ranked probability score (CRPS) is used to assess the quality of the predictive distribution, and is defined as $\mathrm{CRPS} = \frac{1}{M}\sum_{m=1}^{M}|\boldsymbol{\theta}_m-\boldsymbol{\theta}_{true}| - \frac{1}{2M^2}\sum_{m=1}^{M}\sum_{m'=1}^{M}|\boldsymbol{\theta}_m-\boldsymbol{\theta}_{m'}|$, 
where $\boldsymbol{\theta}_m$ and $\boldsymbol{\theta}_{m'}$ denote the samples in the posterior sampling, and $\boldsymbol{\theta}_{true}$ is the true parameter set used to generate the synthetic data. It is a scoring rule that rewards predictive distributions that are both concentrated and well calibrated around an observed value, where smaller CRPS values indicate better predictive performance \citep{gneitingStrictlyProperScoring2007}. True parameter values are known by construction in all simulation experiments, permitting direct evaluation of both point estimation accuracy and posterior calibration.

\subsubsection{Point estimation}
\label{sec:results-point-screening}
 
\begin{figure}[!t]%
  \centering
  \includegraphics[width=1\textwidth, keepaspectratio]{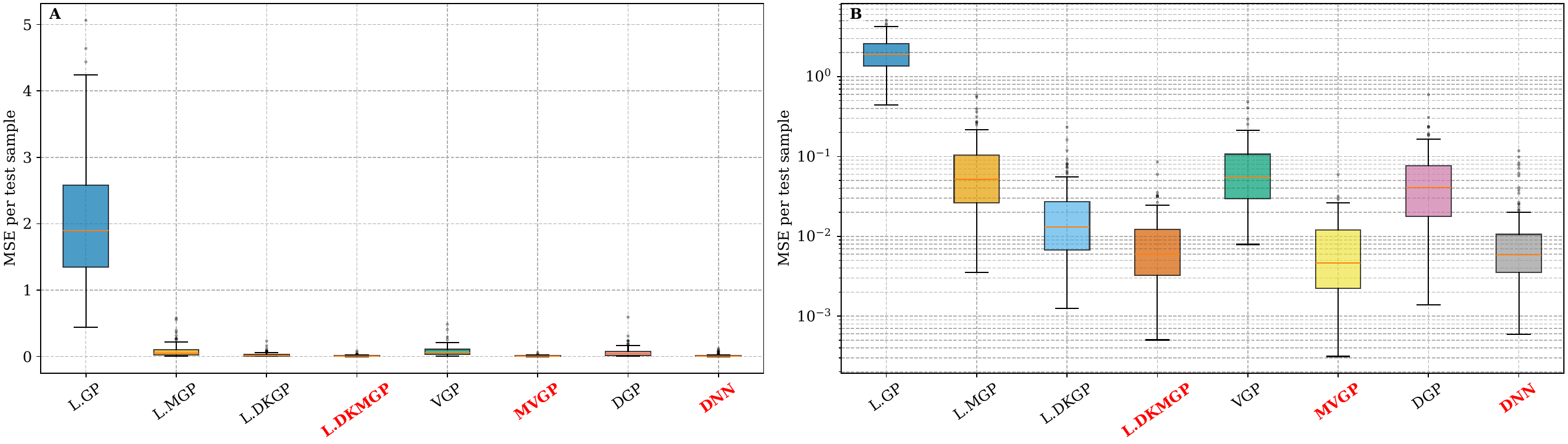}
  \caption{Per-sample MSE distributions for eight emulator architectures evaluated on the test set. Panel (A) shows the distributions on the original scale, emphasising absolute differences in prediction error, whereas panel (B) shows the same distributions on a logarithmic scale, revealing proportional differences across orders of magnitude. In both panels, the models are ordered as L.GP, L.MGP, L.DKGP, L.DKMGP, VGP, MVGP, DGP, and DNN. The three best-performing models (MVGP, L.DKMGP, and DNN) are highlighted in red on the horizontal axis. See Table~\ref{tab:emulator-architectures} for details of the emulator architectures.}
  \label{fig:per_sample_mse}
  \figalttext[Per-sample MSE distributions]{Two side-by-side box-and-whisker plots compare per-test-sample mean squared error across eight emulator architectures. Panel (A) displays MSE per test sample on the original vertical-axis scale, while panel (B) displays the same distributions on a logarithmic vertical-axis scale. In both panels, the horizontal-axis categories are L.GP, L.MGP, L.DKGP, L.DKMGP, VGP, MVGP, DGP, and DNN, with MVGP, L.DKMGP, and DNN highlighted in red as the three best-performing models.}%
\end{figure}

Figure~\ref{fig:per_sample_mse} shows the parameter-estimation MSE across the candidate emulators, multi-output strategies consistently outperformed their single-output counterparts. MVGP achieved the lowest MSE in the parameter space, followed by L.DKMGP and the DNN. Within each pairwise comparison of single- versus multi-output variants (L.GP vs.~L.MGP, L.DKGP vs.~L.DKMGP, and VGP vs.~MVGP), Wilcoxon signed-rank tests with Holm--Bonferroni correction rejected the null hypothesis of equal performance at the $\alpha = 0.05$ level. This pattern is consistent with multi-output strategies that exploit cross-output correlations, which the independent single-output GPs ignore. The single-output L.GP performed worst, indicating that independently trained GPs were not sufficiently expressive for the present regional estimation task.

We also pass back each set of estimated parameters through the forward model to verify that the parameter-space ranking reflected genuine recovery of the inverse mapping rather than coincidental output-space agreement. The full numerical comparison is reported  in Table~S1 of the Supplementary Material. The two rankings were closely aligned: the top three emulators by parameter-space MSE (MVGP, L.DKMGP, DNN) also achieved the lowest output-space MSE, suggesting that these models provided comparatively accurate approximations to the inverse mapping. Notably, L.GP exhibited reasonable output-space accuracy despite poor parameter-space performance, in which distinct parameter configurations produce similar simulator outputs and therefore mask large parameter-space errors. Based on these results, we retain MVGP, L.DKMGP, and the DNN for the next assessment.

\subsubsection{Uncertainty quantification}

Posterior inference for the three emulator implementations retained for uncertainty evaluation was performed using NUTS as described in Section~\ref{sec:bayesian-inference}. Across models and parameters, the median $\widehat{R}$ values ranged from $0.9995$ to $1.0005$, and all the mean effective sample sizes were larger than $400$. These aggregate summaries did not indicate systematic problems with between-chain consistency or sampling efficiency. Full summaries are reported in Tables~S3--S5 of the Supplementary Material.

\begin{figure}[h]%
  \centering
  \includegraphics[width=0.85\textwidth, keepaspectratio]{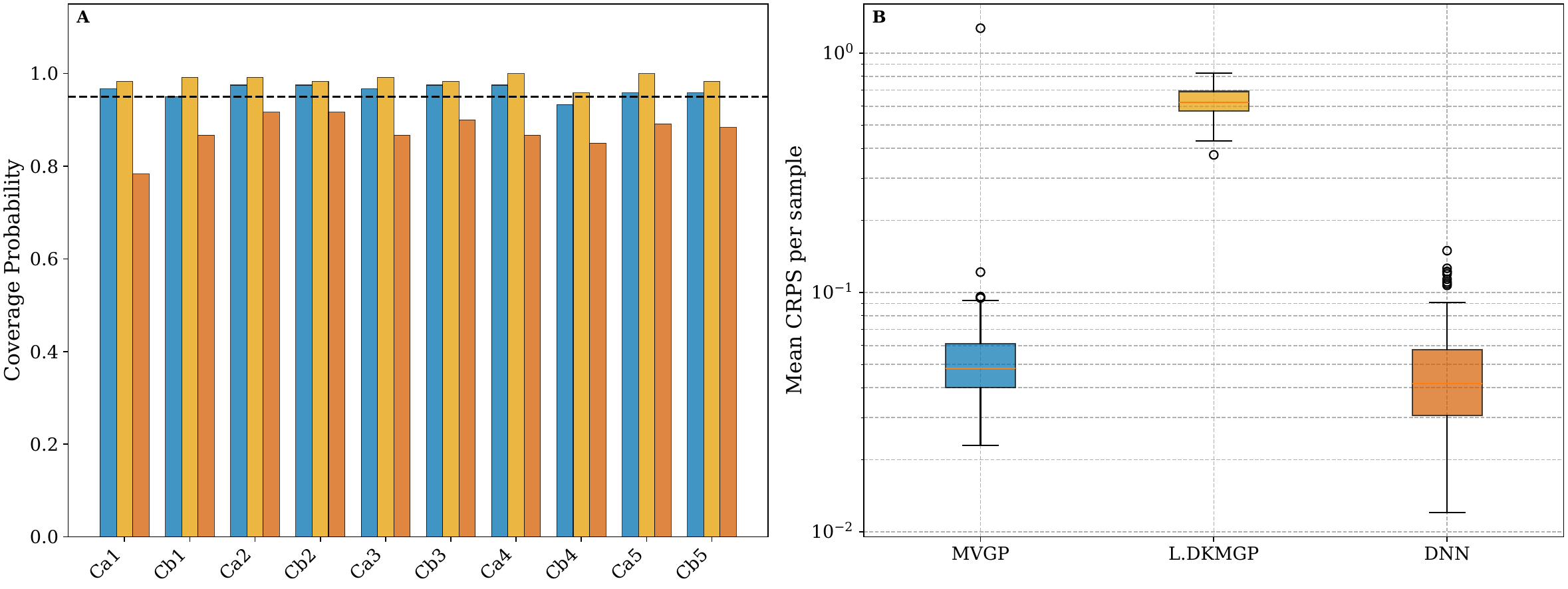}
  \caption{Uncertainty-quantification performance of the three top-ranked emulators: MVGP, L.DKMGP, and DNN. Panel (A) shows the empirical 95\% coverage probability for each parameter. The black dashed horizontal line denotes the nominal coverage level of 0.95. MVGP and L.DKMGP achieve near this coverage across the parameters, whereas DNN exhibits undercoverage. Panel (B) shows the distributions of mean CRPS per test sample on a logarithmic scale. MVGP and DNN have comparably low CRPS values, while L.DKMGP produces substantially higher scores, indicating poorer probabilistic predictive performance despite its near-nominal coverage. MVGP, L.DKMGP, and DNN are represented by blue, orange, and vermillion fills, respectively.}
  \label{fig:per_sample_crps}
  \figalttext[Coverage probabilities and per-sample CRPS distributions]{Two-panel comparison of uncertainty-quantification performance for MVGP, L.DKMGP, and DNN. Panel (A) shows the 95 percent credible-interval coverage probability by parameter, with a black dashed horizontal line marking the nominal value of 0.95. MVGP and L.DKMGP are generally close to the nominal level, whereas DNN shows lower coverage. Panel (B) shows box-and-whisker distributions of mean CRPS per test sample on a logarithmic vertical-axis scale; MVGP and DNN generally have lower CRPS values than L.DKMGP. MVGP is shown in blue, L.DKMGP in orange, and DNN in vermillion.}%
\end{figure}

We next assess posterior uncertainty through the empirical coverage of the marginal $95\%$ credible intervals (CI), as shown in Figure~\ref{fig:per_sample_crps}. Average marginal coverage across the ten parameters was $96.3\%$ for MVGP and $96.6\%$ for L.DKMGP, both close to the reasonable level. The corresponding average for the DNN was $88.1\%$, with coverage for $C_{a1}$ falling to $78.3\%$, indicating that its marginal posterior intervals were generally too concentrated and that posterior uncertainty was underestimated.

\begin{figure}[!t]%
  \centering
  \includegraphics[width=0.85\textwidth, keepaspectratio]{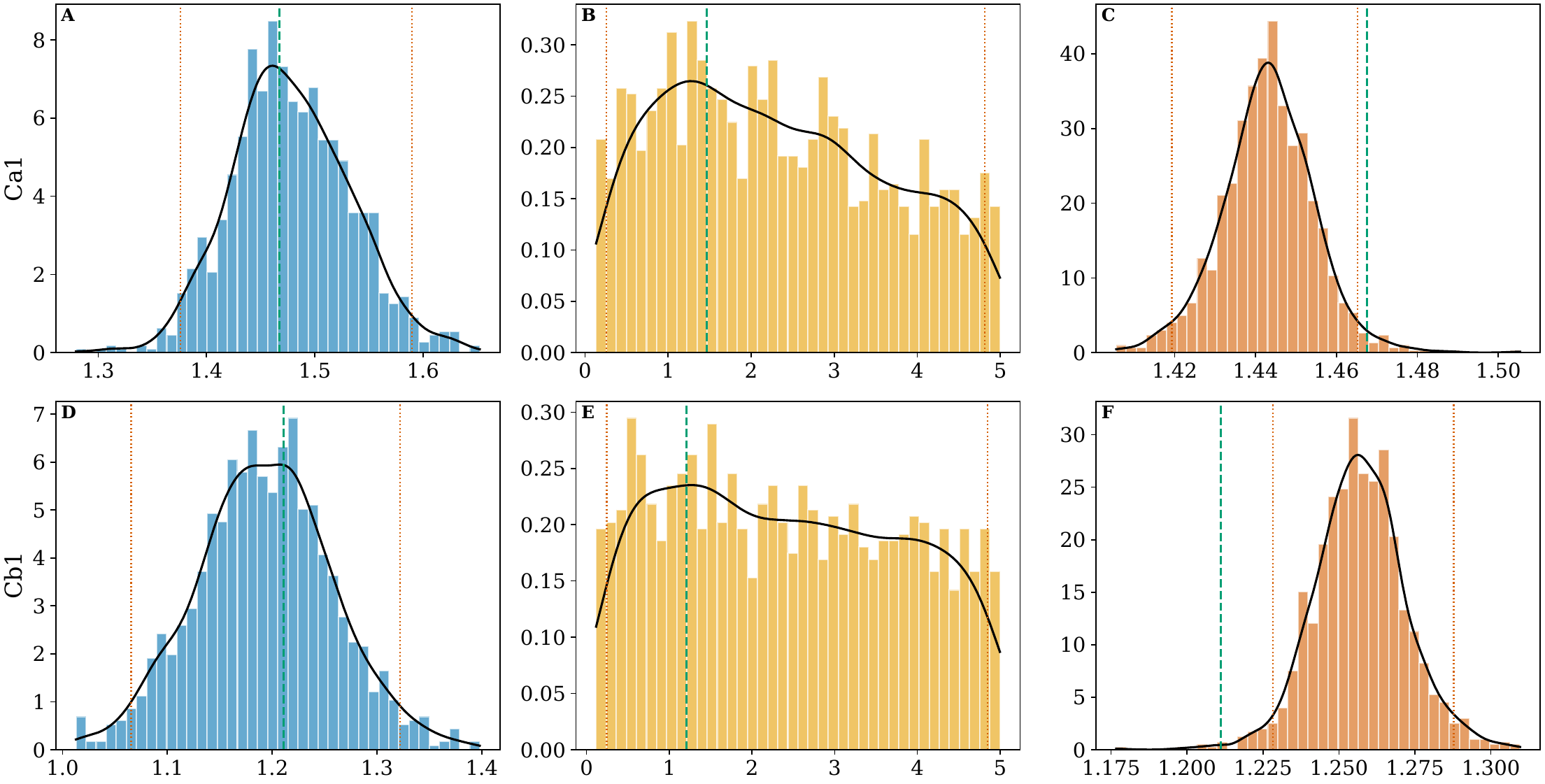}
  \caption{MCMC trace plots and marginal posterior distributions for the representative parameters $C_a^{(1)}$ and $C_b^{(1)}$ in one test case. Panels (A)--(C) show $C_a^{(1)}$ under MVGP, L.DKMGP, and DNN, respectively; panels (D)--(F) show the corresponding results for $C_b^{(1)}$. Green dashed lines indicate the generating parameter values, and vermillion dotted lines indicate the bounds of the $95\%$ credible intervals.}
  \label{fig:trace_posterior}
  \figalttext[MCMC traces and posterior distributions for two representative parameters]{Six-panel figure arranged in two rows and three columns. The top row, panels A to C, shows MCMC traces and marginal posterior distributions for the stiffness-magnitude parameter Ca1 under MVGP, L.DKMGP, and DNN. The bottom row, panels D to F, shows the corresponding results for the nonlinear-stiffening parameter Cb1. Green dashed lines mark the generating values, and vermillion dotted lines mark the 95 percent credible-interval bounds.}%
\end{figure}

Figure~\ref{fig:per_sample_crps} shows that the mean marginal CRPS produced a different ordering from the coverage results, with L.DKMGP giving the highest value and DNN achieving a slightly lower value than MVGP. Although this ranking would favour DNN if CRPS were considered in isolation, the marginal posterior distributions in Figure~\ref{fig:trace_posterior} display excessively narrow intervals that failed to contain the truth, so the apparent CRPS advantage was driven by over-confidence. L.DKMGP showed the opposite pattern, combining coverage of over $95\%$ with comparatively diffuse marginal posteriors and weak posterior contraction. These results show that CRPS should be interpreted jointly with coverage and posterior contraction rather than used as a stand-alone model-selection criterion. The complete posterior trajectories and marginal distributions are presented in Figure~S1 of the Supplementary Material.

Posterior contraction was quantified using the posterior-to-prior variance ratio $\mathrm{VR}_d = \frac{\mathrm{Var}_{\mathrm{post}}(\theta_d)}{\mathrm{Var}_{\mathrm{prior}}(\theta_d)}$ computed separately for each of the ten regional parameters, where lower values indicate a greater reduction in parameter uncertainty relative to the prior. Parameter-specific results are reported in Table~S2 of the Supplementary Material. All ten parameter-specific mean ratios for MVGP were below $0.04$, indicating substantial posterior contraction, whereas L.DKMGP retained most of the prior variance and therefore provided comparatively limited information about the regional parameters. Although the DNN produced the strongest contraction, its lower empirical coverage indicates that the corresponding credible intervals were typically too narrow rather than more reliable. In every zone and for all three implementations, the $C_a^{(z)}$ parameters exhibited lower variance ratios than the paired $C_b^{(z)}$ parameters. Under the end-diastolic observation design, the stiffness-magnitude parameters were therefore more strongly constrained by the data than the parameters governing nonlinear stiffening.

Considered together with the point-estimation, marginal-coverage and marginal-CRPS results, these findings support the selection of MVGP, which provided the most favourable balance between parameter-recovery accuracy, posterior coverage and contraction. Unless stated otherwise, all subsequent inference results were obtained using MVGP.

\subsection{Inference under simulated regional material stiffness}
\label{sec:results-stiffness}
 
We applied the MVGP-based inference framework to the four simulated scenarios introduced in Section~\ref{sec:simulated-disease-scenarios}. The following subsections examine discrimination between stiffened and baseline regions, posterior interval widths, and posterior dependence.

\subsubsection{Distinguishing stiffened from baseline regions}
\label{sec:results-distinguishing}

\begin{figure}[!t]
  \centering
  \includegraphics[width=0.85\textwidth, keepaspectratio]{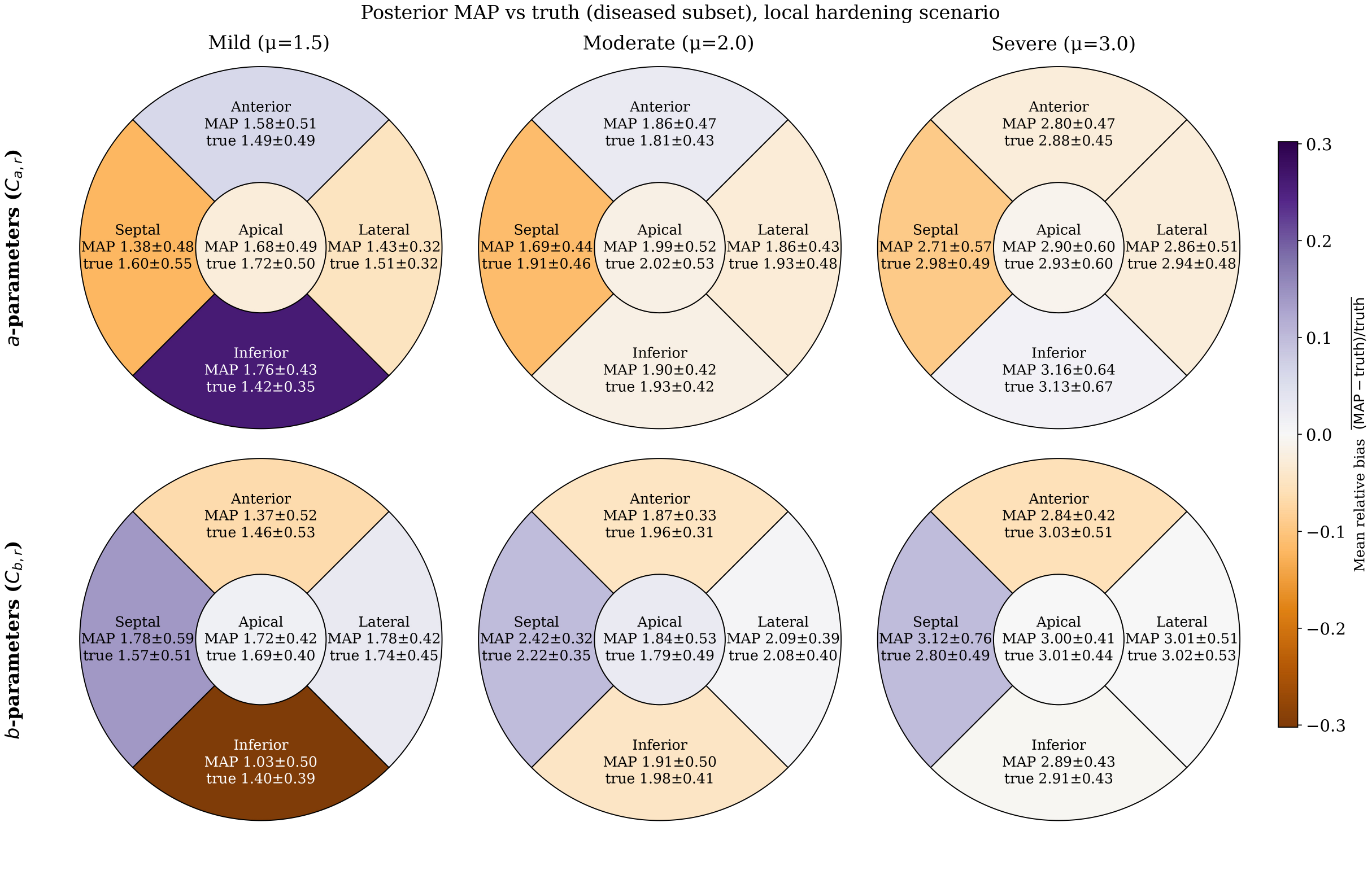}
  \caption{Bullseye comparison of MAP estimates against the imposed ground-truth values in the stiffened zone, for the three local stiffening scenarios ($\mu \in \{1.5, 2.0, 3.0\}$). This figure presents only the results from the stiffening regions in the local stiffening scenarios, the results for the baseline regions are shown in Figure \ref{fig:bullseye_health_truth_vs_inferred}.}
  \label{fig:bullseye_truth_vs_inferred}
  \figalttext[Bullseye plots, local stiffening]{Bullseye comparison of inferred and ground-truth regional parameters under local stiffening at three severity levels.}
\end{figure}

The principal diagnostic question is whether the inferred parameter posteriors can distinguish stiffened regions from healthy myocardium. Figure~\ref{fig:bullseye_truth_vs_inferred} provides the first of three complementary visualisations used to examine this question by comparing the maximum a posteriori (MAP) estimates of $C_a^{(z)}$ and $C_b^{(z)}$ with their generating values in the stiffened zone of each focal scenario.

Figure~\ref{fig:bullseye_truth_vs_inferred} reports the MAP estimates of $C_a$ and $C_b$ in the stiffened zone of each local stiffening scenario, displayed on the AHA bullseye against the ground-truth values; for each anatomical zone, the displayed sample is taken from the local scenarios in which that zone is the designated stiffened target. Under severe local stiffening ($\mu = 3.0$), the parameter shift is the largest, and the MAP estimates align most closely with the imposed truth, producing the sharpest separation between the inferred values and the baseline cluster. The moderate scenario, with $\mu=2.0$, shows a middle pattern in which the estimated regional change remains visibly separated from the reference value, although the estimation errors are larger than under severe stiffening. Under mild local stiffening ($\mu = 1.5$), where the shift approaches the baseline noise level, the MAP errors are larger. However, the inferred values nonetheless remain distinguishable from the baseline, indicating that the framework can identify regional stiffening even at modest severities, albeit with reduced precision.

\begin{figure}[!t]
  \centering
  \includegraphics[width=0.65\textwidth, keepaspectratio]{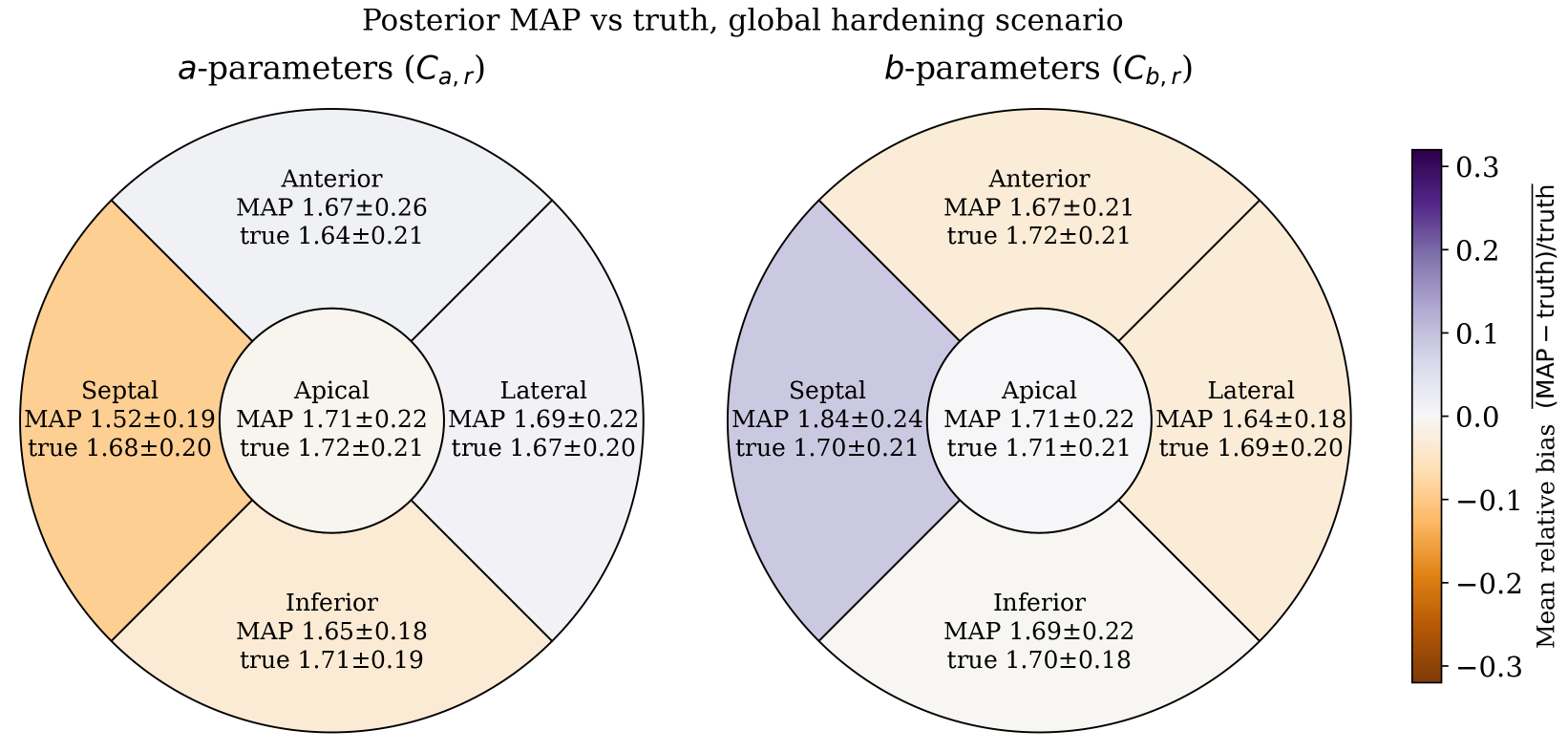}
  \caption{Bullseye comparison of MAP estimates against the imposed ground-truth values for the multi-territory stiffening scenario ($\mu = 1.7$). All five zones are simultaneously perturbed; the inferred MAP and the imposed truth are shown for $C_a$~(left) and $C_b$ (right). MAP-truth agreement is consistently good across all zones and is on average closer than under local stiffening at any severity.}
  \label{fig:bullseye_HCM_truth_vs_inferred}
  \figalttext[Bullseye plots, multi-territory stiffening]{Bullseye comparison of inferred and ground-truth regional parameters under multi-territory stiffening.}
\end{figure}

Figure~\ref{fig:bullseye_HCM_truth_vs_inferred} reports the analogous comparison for the global stiffening scenario. The MAP errors are systematically smaller than those obtained under local stiffening at any of the three severities, and they are distributed approximately uniformly across the five zones rather than concentrated in any one territory. Although a simultaneous perturbation of every zone might be expected to be more challenging, the consistent shift in all regions produces a more identifiable global strain signal than a perturbation confined to a single territory, with the apparent additional difficulty manifesting instead through posterior uncertainty rather than through MAP bias, as discussed in Sections~\ref{sec:results-hpd}--\ref{sec:results-correlation}.

\begin{figure}[!t]
  \centering
  \includegraphics[width=0.85\textwidth, keepaspectratio]{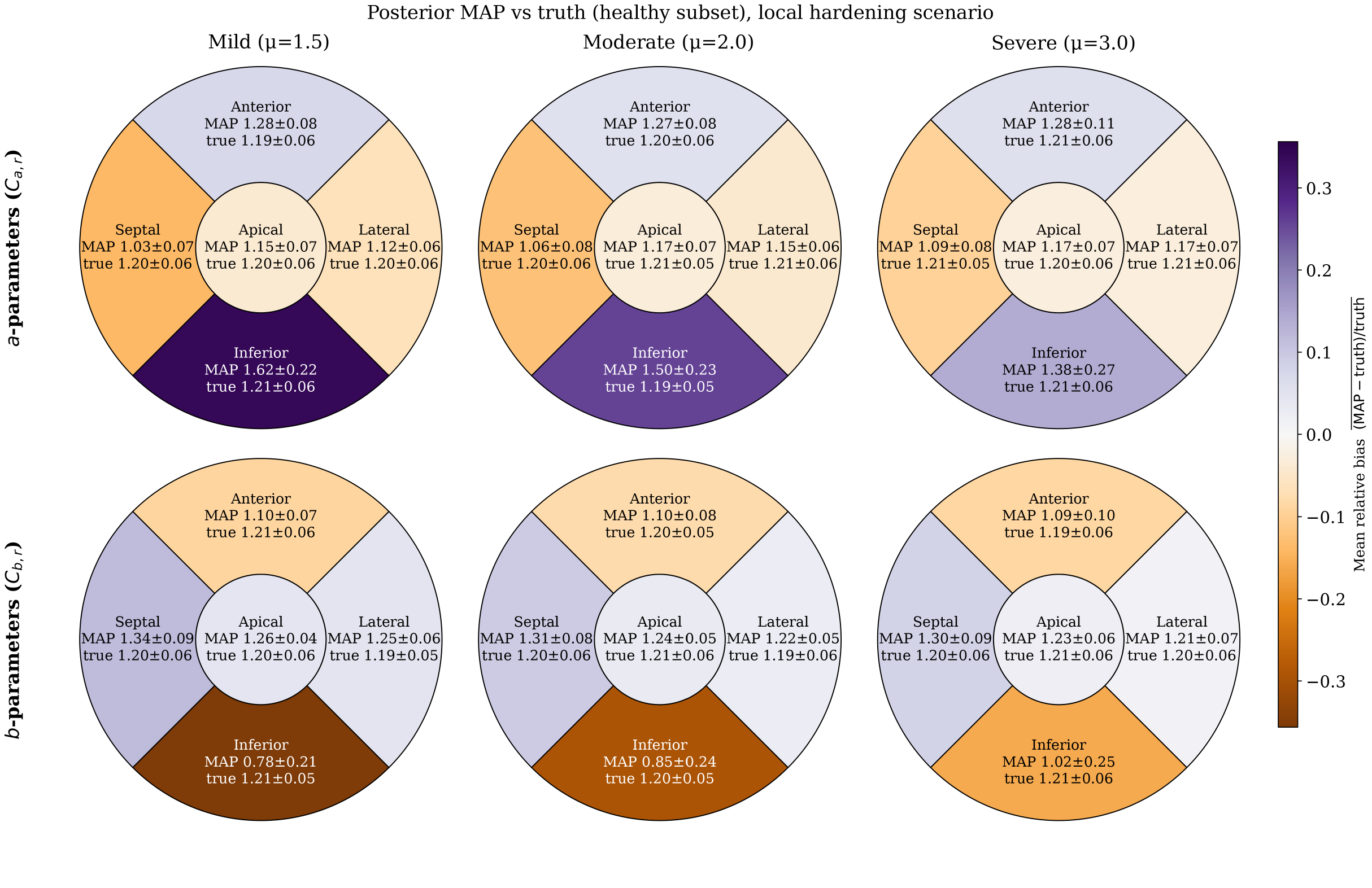}
  \caption{Bullseye comparison of MAP estimates against the imposed ground-truth values for the baseline (non-stiffened) regions, aggregated across all four stiffening scenarios. The inferred MAP and the imposed truth are shown for $C_a$ (top row) and $C_b$ (bottom row). MAP estimates concentrate near $C_a = C_b = 1$ and are clearly displaced from the stiffened-zone estimates in Figures~\ref{fig:bullseye_truth_vs_inferred} and~\ref{fig:bullseye_HCM_truth_vs_inferred}.}
  \label{fig:bullseye_health_truth_vs_inferred}
  \figalttext[Bullseye plots, baseline regions]{Bullseye comparison of inferred and ground-truth regional parameters for baseline regions across all stiffening scenarios.}
\end{figure}

Figure~\ref{fig:bullseye_health_truth_vs_inferred} reports MAP estimates for the baseline regions, aggregated across all four scenarios. The MAP estimates concentrate near $C_a = C_b = 1$ and are visibly distinct from the corresponding stiffened-zone estimates in Figures~\ref{fig:bullseye_truth_vs_inferred} and~\ref{fig:bullseye_HCM_truth_vs_inferred}. Taken together, the three bullseye comparisons demonstrate that the framework distinguishes stiffened myocardium from healthy regions on the basis of MAP point estimates across all scenarios studied here.

Beyond the stiffened-versus-baseline contrast, the three bullseye figures together expose a regional asymmetry in the spatial distribution of MAP errors. Errors under global stiffening are approximately uniform across the five zones, whereas errors under local stiffening concentrate in the Inferior and Septal walls. The Inferior wall consistently exhibits the largest MAP errors regardless of whether the stiffening is global or local. A potential  interpretation links this region-specific difficulty to the geometric coupling of the Inferior wall in the specific LV geometry studied here (e.g. thinner wall thickness and higher curvature compared to the Septum), rather than to a limitation of the inference procedure itself, which is discussed further in Section~\ref{sec:Discussion}.

\subsubsection{Posterior uncertainty and HPD interval widths}
\label{sec:results-hpd}

\begin{figure}[!t]
  \centering
  \includegraphics[width=0.85\textwidth, keepaspectratio]{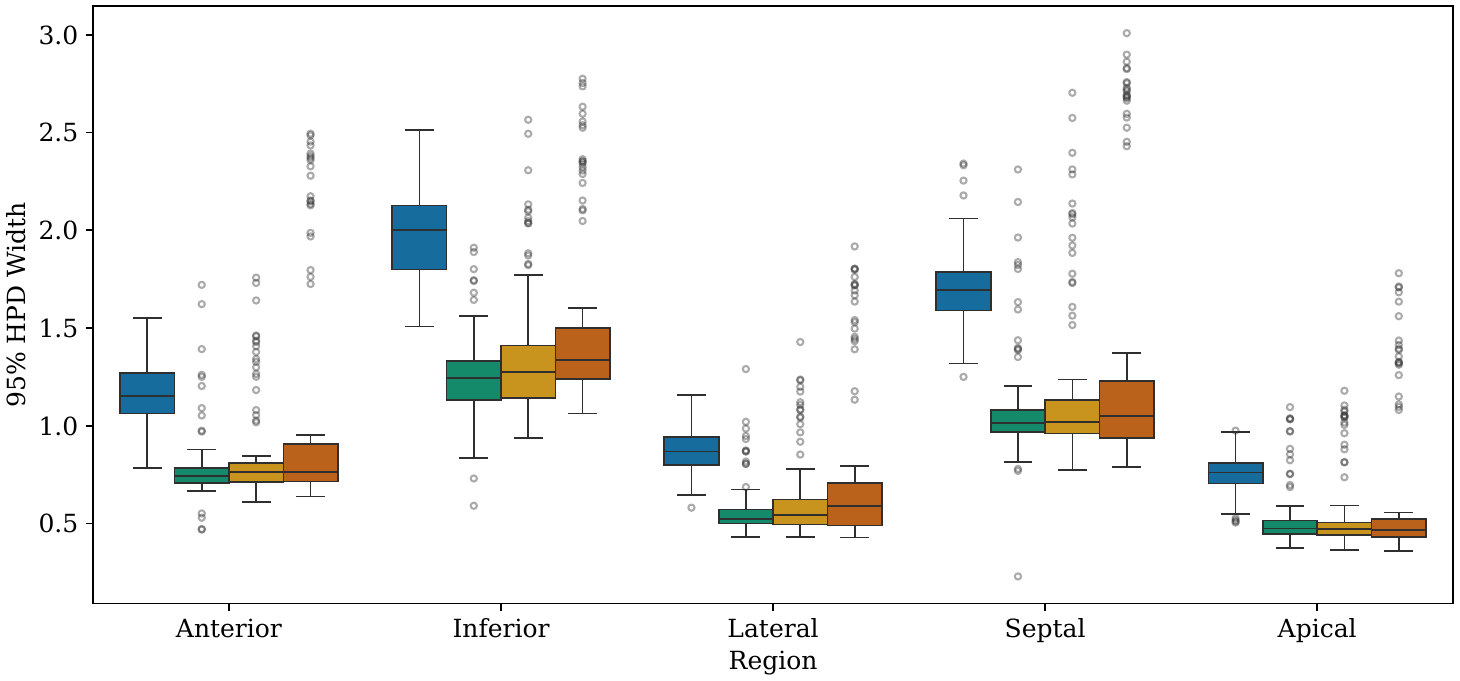}
  \caption{Box plots of $95\%$ HPD interval widths across the four stiffening scenarios. Each box summarises the HPD widths of the ten regional parameters over all test samples within a given scenario; broader boxes indicate greater posterior conservatism. HPD widths are systematically broader under multi-territory stiffening than under local stiffening, and tend to grow with severity within the local scenarios.}
  \label{fig:HPD}
  \figalttext[HPD widths box plots]{Box plots of 95 percent HPD interval widths across four stiffening scenarios.}
\end{figure}

Whilst the MAP estimates above demonstrate point-level discrimination, the Bayesian framework additionally quantifies posterior uncertainty. Figure~\ref{fig:HPD} shows box plots of the $95\%$ highest posterior density (HPD) interval widths across the four scenarios. Posterior uncertainty is generally conservative, meaning HPD intervals are wider than the spread of MAP estimates. However, the MAP errors reported in Section~\ref{sec:results-distinguishing} remain low, indicating that the true parameter values lie close to the posterior mode. The conservatism, therefore, does not compromise the accuracy of the inference, although it does limit the precision with which any single regional parameter can be constrained from one observation.

The HPD widths nonetheless tend to grow with severity, particularly in the stiffened region itself. A more pronounced regional contrast amplifies posterior uncertainty for the stiffened parameters, whilst simultaneously sharpening inference in the surrounding baseline regions, whose strain values become easier to estimate against the perturbation.

\subsubsection{Posterior correlation structure}
\label{sec:results-correlation}

The conservatism of the posterior intervals and the high inference errors in the Inferior and Septal walls may both be related to distinct geometrical features among regional parameters in LV mechanics. If an increase in the parameter of one region is partially offset by a decrease in the parameter from another region, the resulting strain pattern can be nearly identical, and such compensatory mechanisms manifest in the posterior as structured pairwise correlations. To characterise this structure, we extracted the parameter posterior correlations from each MCMC sample and averaged them across test samples within each scenario.

\begin{figure}[!t]
  \centering
  \includegraphics[width=0.85\textwidth, keepaspectratio]{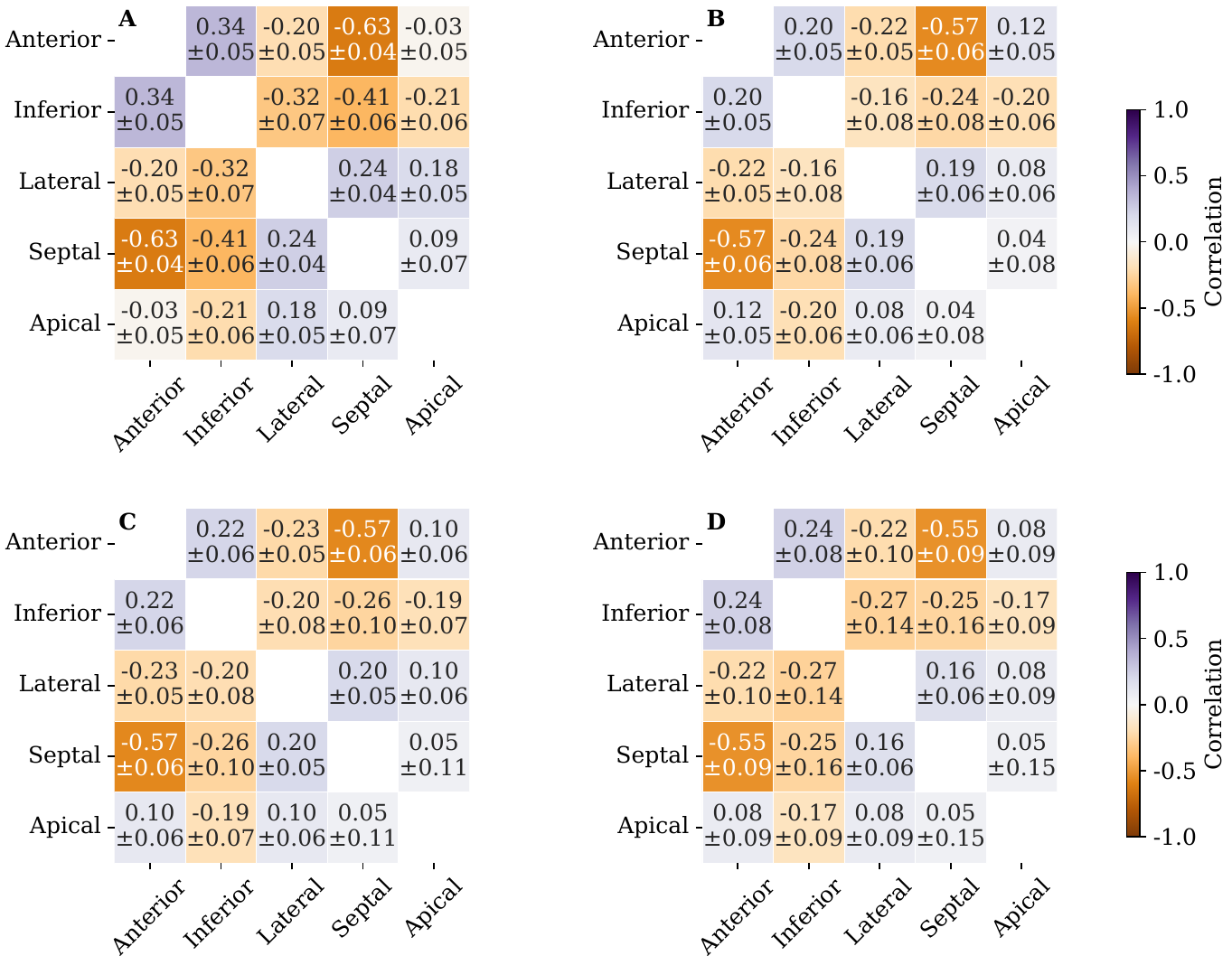}
  \caption{Inter-regional posterior correlation matrices obtained by first-principal-component projection within each region, across the four stiffening scenarios. Each cell displays the mean correlation $\pm$ one standard error aggregated over test samples. The Anterior--Septal pair exhibits a strong negative correlation that persists across all four scenarios, whilst additional inter-regional couplings (Inferior--Septal, Anterior--Inferior, and Inferior--Lateral) emerge only under multi-territory stiffening.}
  \label{fig:inter_region_correlation}
  \figalttext[Inter-regional correlation matrices]{Heatmap matrices showing inter-regional posterior correlations via PC1 projection across four stiffening scenarios.}
\end{figure}

Within each region, $C_a$ and $C_b$ exhibited a near-deterministic negative correlation across all four scenarios; full intra-regional details are reported in Figure~S2-S5 of the Supplementary Material. This intra-regional coupling reflects the inherent mechanical trade-off between stiffness magnitude and the rate of nonlinear stiffening within a given territory: increasing one and decreasing the other can produce nearly identical regional strain at end diastole. Motivated by this strong intra-regional structure, we projected the $10 \times 10$ posterior correlation matrix onto its first principal component within each region, reducing the inter-regional structure to a $5 \times 5$ matrix. The resulting matrices for all four scenarios are presented in Figure~\ref{fig:inter_region_correlation}.

The inter-regional correlation pattern depends more strongly on the spatial pattern of stiffening than on its severity. Under multi-territory stiffening, four stable pairwise couplings emerged: Anterior--Septal ($\bar{r} \approx -0.63$), Inferior--Septal, Anterior--Inferior, and Inferior--Lateral. Under all three local stiffening severities, however, only the negative correlation of the neighbouring regions Anterior--Septal persisted, ranging from $-0.55$ to $-0.57$, with all other inter-regional couplings attenuating substantially. The Apical zone showed no appreciable inter-regional correlations under any scenario, an absence that may be associated with the smaller physiological strain typically observed at the apex relative to the basal and mid-cavity zones.

\subsection{Clinical data application}
\label{sec:Clinical data application}

\begin{figure}[!t]
  \centering
  \includegraphics[width=0.9\textwidth, keepaspectratio]{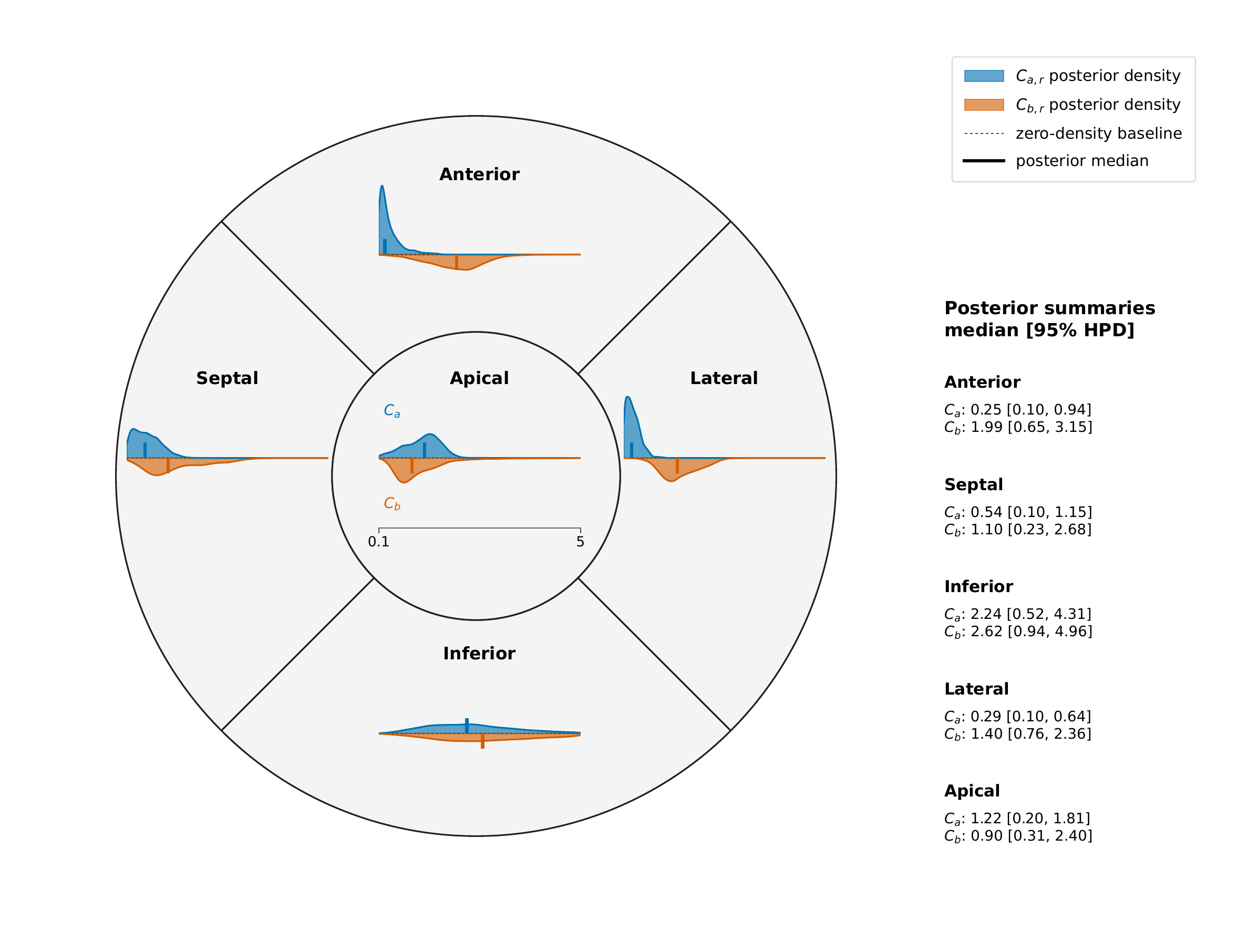}
  \caption{Posterior distributions of the regional H-O parameters for the clinical case, displayed on the five-zone AHA bullseye. Within each zone, the upper blue density represents $C_a^{(z)}$ and the lower vermillion density represents $C_b^{(z)}$. The grey dotted line is the zero-density baseline, and the short vertical markers indicate the posterior medians. Numerical summaries are reported as posterior median [$95\%$ HPD interval]. The inferred direction-specific noise scales were $\sigma_1=0.353\,[0.302,\,0.409]$ for circumferential strain and $\sigma_2=0.393\,[0.343,\,0.453]$ for radial strain.}
  \label{fig:bullseye_posterior_realcase}
  \figalttext[Bullseye posterior for clinical case]{Five-zone AHA bullseye containing paired posterior densities for the regional H-O parameters. In each zone, the blue density for Ca is displayed above a grey dotted zero-density baseline and the vermillion density for Cb below it; short vertical markers indicate posterior medians. Numerical summaries give the posterior median and 95 percent HPD interval for each parameter and for the circumferential and radial noise scales.}
\end{figure}

The MVGP framework was applied to the healthy-volunteer observations described in Section~\ref{sec:clinical-data}. A separate MVGP was trained on the 33 simulator outputs corresponding to the available clinical measurements. At this lower output dimension, the predictive-covariance instability that motivated the principal-component representation in Section~\ref{sec:results-pca} was not observed, and applying PCA reduced predictive accuracy relative to modelling the retained outputs directly. We therefore used the original output representation, preserving the correspondence between emulator outputs and the observed strain channels and allowing direction-specific measurement-error scales to be defined in the observation space. The resulting 12-dimensional posterior was sampled using the NUTS framework of Section~\ref{sec:bayesian-inference}. Figure~S6 in the Supplementary Material presents the corresponding trace plots and marginal posterior densities, while Figure~\ref{fig:bullseye_posterior_realcase} summarises the regional posterior using posterior median and $95\%$ HPD interval widths, with the inferred noise scales reported alongside.

\begin{figure}[!t]%
  \centering
  \includegraphics[width=0.9\textwidth]{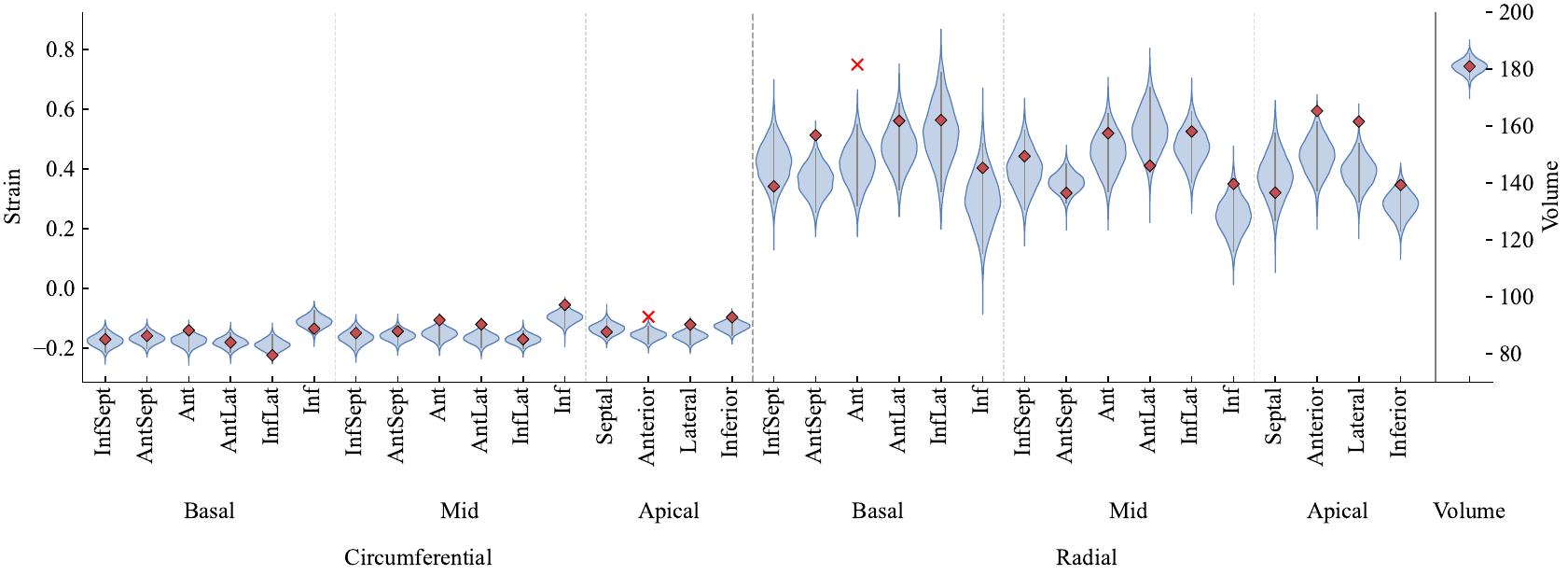}
  \caption{Posterior predictive checks for the clinical case. The violin outlines are kernel density estimates of the posterior predictive distributions for each of the 32 regional strain outputs and the volume. Strain is read against the left-hand axis; volume is read against the right-hand axis. The red diamonds denote observed values that lie within the 95\% credible interval of the corresponding predictive distribution and the red crosses denote observed values that lie outside. Two of the 33 observations fall outside.}
  \label{fig:posterior_predictive_clinical}
  \figalttext[Posterior predictive violins]{A single wide row of violin plots. The circumferential violins are narrow and lie entirely below zero; the radial violins lie entirely above zero and are roughly four times taller, their shapes varying more from segment to segment. A separate narrow violin stands apart at the far right. Most red markers sit near the widest part of their violin, whereas the two crosses sit clear of the outline, one among the apical circumferential violins and one among the basal radial violins.}
\end{figure}

The results shown in Figure~\ref{fig:bullseye_posterior_realcase} indicate that the posterior distributions for most parameters have contracted significantly, however, parameter uncertainty is greatest in the Inferior region, a finding consistent with the results obtained from previous simulation data. To assess the resulting inference, we constructed posterior predictive distributions for each output dimension. Each posterior parameter sample was passed through the MVGP emulator and, by the nature of the GP, every such forward pass yields a predictive distribution rather than a point. Samples were drawn from this predictive distribution, and the sampled measurement-noise variance was added, producing a posterior predictive interval that jointly accounts for parameter uncertainty, emulator predictive uncertainty, and observational noise. The resulting predictive distributions are shown in Figure~\ref{fig:posterior_predictive_clinical}.

The posterior predictive distributions covered most of the healthy-volunteer observations. The end-diastolic volume lay near the centre of its predictive distribution, while only two of the $32$ strain observations fell outside their corresponding $95\%$ posterior predictive intervals. These results indicate that the fitted model reproduced most of the observed volume and strain measurements for this individual.

Posterior predictive uncertainty was greater for radial than for circumferential strain, with wider radial intervals and a larger posterior estimate for the radial noise scale, $\sigma_2$, than for the circumferential noise scale, $\sigma_1$, which is consistent with the imaging literature \citep{gaoLeftVentricularStrain2014}. Overall, the clinical analysis demonstrates the feasibility of jointly inferring regional constitutive parameters and direction-specific measurement uncertainty from in vivo observations, while further subject-level validation is required before the results can be interpreted clinically.

\newpage
\section{Discussion}
\label{sec:Discussion}

The proposed framework combines a five-zone H-O parameterisation with emulator-based Bayesian inference for regional LV mechanics, and the evaluations address three questions. The first is which emulator strategies support accurate recovery of the ten regional parameters under a fixed simulation budget, and whether point-estimation accuracy alone is sufficient for deciding upon an emulator for posterior inference. The second is what the end-diastolic observation design reveals about the regional stiffness-magnitude and nonlinear-stiffening parameters. The third is whether inference remains feasible when the clinical observation vector is incomplete, and the strain-noise scales are unknown. Because all analyses use a single fixed LV geometry and the clinical application involves one healthy volunteer, generalising across ventricular geometries and patient populations will require validation on additional clinical datasets.

The emulator comparison in Section~\ref{sec:results-emulator-selection} shows that surrogates that exhibit comparable point-estimation performance can still result in posterior distributions that differ substantially with respect to marginal coverage and contraction. Among the candidate implementations, MVGP provided the most favourable balance between parameter recovery, marginal coverage and posterior contraction. The undercoverage observed for the DNN suggests that the single shared error scale used with its deterministic predictions did not fully represent uncertainty in this setting, which may require the evaluation of more flexible uncertainty formulations for neural-network emulators in the future. The weak contraction observed for L.DKMGP indicates that the likelihood induced by its local predictive distributions provided comparatively limited discrimination among parameter values. Generally, the findings support assessing an emulator using criteria relevant to the inference end goal, rather than predictive error alone.

The variance-ratio results in Section~\ref{sec:results-emulator-selection} indicate that the end-diastolic observations provided more information about the stiffness-magnitude parameters $C_a^{(z)}$ than about the nonlinear-stiffening parameters $C_b^{(z)}$. Under the reduced H--O parameterisation, $C_a^{(z)}$ controls the overall material-stiffness scale in zone $z$, whereas $C_b^{(z)}$ controls how rapidly the material response stiffens as deformation increases. Measurements at a single end-diastolic loading state sample only one part of the nonlinear pressure--deformation relationship and therefore provide limited information about the nonlinear stiffening governed by $C_b^{(z)}$. Different combinations of $C_a^{(z)}$ and $C_b^{(z)}$ within a zone can consequently produce similar strain and volume outputs. Accordingly, the observed contrast indicates weaker practical identifiability of $C_b^{(z)}$ from the present end-diastolic observations, while its identifiability under richer observation designs remains an open question. Data spanning multiple loading states, including strain and volume over the diastolic filling trajectory together with pressure information, may better distinguish stiffness magnitude from nonlinear stiffening. This interpretation is consistent with previous sensitivity analyses showing that information about the nonlinear parameters increases at higher loading pressures \citep{lazarusSensitivityAnalysisInverse2022}.

The posterior correlations reported in Section~\ref{sec:results-correlation} provide information about compensatory directions that is not available from marginal posterior summaries alone. In particular, the negative dependence between $C_a^{(z)}$ and $C_b^{(z)}$ within zones and the anterior--septal dependence indicate directions along which parameter changes can compensate under the present simulator, prior, geometry and observation design. These posterior patterns provide exploratory hypotheses about compensatory parameter directions that can be examined through subsequent biomechanical investigation and formal identifiability analysis. The scenario-specific summaries provide further context: the global stiffening scenario exhibited several inter-zone dependencies, whereas the anterior--septal negative dependence was the only inter-zone pattern that persisted across all three local severities. When considered alongside the HPD-width and MAP comparisons in Sections~\ref{sec:results-stiffness}, the global scenario combined broader HPD intervals with comparatively well-centred MAP estimates, while the local scenarios showed narrower intervals but larger zone-specific MAP deviations. One possible interpretation is that a spatially coherent global strain signal helps to locate the posterior mode even when several compensatory directions remain, whereas a local strain signal is more susceptible to ambiguity between neighbouring zones. This interpretation is descriptive rather than causal, because the present experiments do not isolate how posterior dependence affects interval width or estimation error, and identifying the underlying mechanism would require a structured sensitivity analysis and replication across ventricular geometries.

The clinical application handled the incomplete observation vector by using the corresponding subset of the simulator output, and direction-specific strain-noise scales can be estimated jointly with the constitutive parameters. Joint estimation of the noise scales yielded posterior predictive distributions that reproduced most of the available observations, providing an internal check on the fitted model. This result establishes the feasibility of jointly inferring regional material parameters and direction-specific noise scales from the available observations. Extending the analysis to multiple subjects, ventricular geometries and repeated measurements will enable the reproducibility and transportability of both the directional noise pattern and the broader inference framework to be assessed across patient populations and imaging protocols. This is discussed in Section~\ref{sec:Future work}.

\section{Future work}
\label{sec:Future work}

The most immediate extension is to generalise the emulator beyond the single LV geometry considered here. \citet{lazarusImprovingCardioMechanicInference2022} developed a multi-geometry emulator in which LV anatomy was represented by five principal-component scores and combined with four material-property parameters to predict end-diastolic cavity volume and circumferential strains. This construction provides a direct template for defining a low-dimensional geometry descriptor $\mathbf{g}$ and augmenting the emulator input from $\boldsymbol{\theta}$ to $(\boldsymbol{\theta},\mathbf{g})$. The regional setting considered here would extend that construction to ten material parameters and a higher-dimensional, multi-directional segmental-strain output. Training over the resulting joint input space would support inference across patient-specific anatomies and allow the anterior--septal dependence and the relatively large inferior-zone MAP errors reported in Sections~\ref{sec:results-correlation} and~\ref{sec:results-distinguishing}, respectively, to be examined across geometries. The principal design challenge would be to cover both regional material parameters and anatomical variation within a finite simulation budget. The Sobol design may then provide insufficient local resolution, motivating structured designs that allocate simulations preferentially to regions of high emulator uncertainty or posterior relevance in $(\boldsymbol{\theta},\mathbf{g})$.

A further extension is to replace the end-diastolic data with observations collected throughout diastolic filling. Time dependent strain and volume measurements may improve the separation of the stiffness-magnitude parameters $C_a^{(z)}$ from the nonlinear-stiffening parameters $C_b^{(z)}$ by probing the material response at multiple loading states. \citet{geAdvancedStatisticalInference2025} developed a time-series Gaussian process emulator for LV cavity volume and incorporated PCA representations of patient-specific geometry, showing that both temporal and anatomical structure can be accommodated within a single emulator. However, volume was the only quantity of interest in that framework, and the method therefore did not address the strain-based regional parameter inference considered here. Extending the present framework would require a joint representation of time-indexed volume and strain indexed by anatomical segment, rather than treating additional time points as an unstructured collection of outputs. This extension would motivate covariance or latent-output constructions that exploit temporal, spatial and cross-output dependence while propagating the resulting emulator uncertainty into the parameter posterior distribution.

\begin{table}[!b]
\caption{Candidate emulator architectures compared in this study. The emulators are the local Gaussian process (L.GP), local multi-output Gaussian process (L.MGP), local deep-kernel Gaussian process (L.DKGP), local deep-kernel multi-output Gaussian process (L.DKMGP), variational Gaussian process (VGP), multi-output variational Gaussian process (MVGP), deep Gaussian process (DGP) and deep neural network (DNN). The characteristics of each method are summarised in the columns.}
\label{tab:emulator-architectures}
\begin{tabular*}{\columnwidth}{@{\extracolsep\fill}lcccc@{}}
\toprule
Emulator & Base method & Training scope & Output structure & Deep component \\
\midrule
L.GP & GP & Local (L) & Single-output & None \\
L.MGP & GP & Local (L) & Multi-output (M) & None \\
L.DKGP & GP & Local (L) & Single-output & Deep kernel (DK) \\
L.DKMGP & GP & Local (L) & Multi-output (M) & Deep kernel (DK) \\
VGP & VGP (V) & Global & Single-output & None \\
MVGP & VGP (V) & Global & Multi-output (M) & None \\
DGP & VGP (V) & Global & Multi-output (M) & VGP layers (D) \\
DNN & NN & Global & Multi-output (M) & Feed-forward network (D) \\
\bottomrule
\end{tabular*}
\end{table}

\section*{Code availability}
The software and source code developed and used in this study are publicly available at \url{https://github.com/Aether-Ren/MultiR_FixedG_JRSSC}.

\section*{Acknowledgments}
VD and HG were supported by EPSRC project EP/Z531182/1. HG further acknowledges the support from EPSRC (EP/T017899/1) and the British Heart Foundation (PG/22/10930). ChatGPT (OpenAI) and Grammarly were used to correct errors of grammar, spelling, punctuation and tone, to improve readability and formatting. These tools were not used to generate original scientific content or results, and all editorial decisions and final wording were reviewed and approved by the authors.

\bibliographystyle{oup-abbrvnat}
\bibliography{oup-authoring-template/references}

\clearpage
\setcounter{section}{0}
\setcounter{figure}{0}
\setcounter{table}{0}
\setcounter{equation}{0}
\renewcommand{\thesection}{S\arabic{section}}
\renewcommand{\thesubsection}{\thesection.\arabic{subsection}}
\renewcommand{\thefigure}{S\arabic{figure}}
\renewcommand{\thetable}{S\arabic{table}}
\renewcommand{\theequation}{S\arabic{equation}}
\renewcommand{\theHsection}{S\arabic{section}}
\renewcommand{\theHfigure}{S\arabic{figure}}
\renewcommand{\theHtable}{S\arabic{table}}
\renewcommand{\theHequation}{S\arabic{equation}}

\setlength{\emergencystretch}{2em}
\section{Emulator selection: numerical comparison}
\label{sec:supp-emulator-selection}

Table~\ref{tab:supp-emulator-ranking} provides the complete numerical comparison of the eight emulator architectures cited in the main manuscript.

\begin{table}[htbp]
\centering
\caption{Numerical ranking of the eight emulator architectures. Param MSE denotes the global mean squared error in parameter space, and Output MSE denotes the mean squared error in output (strain) space. Bold entries identify the three models shortlisted by Param MSE.}
\label{tab:supp-emulator-ranking}
\begin{tabular}{@{}lcc@{}}
\toprule
Model & Param MSE & Output MSE \\
\midrule
L.GP    & 2.0639 & 0.003194 \\
L.MGP   & 0.0853 & 0.009780 \\
L.DKGP  & 0.0239 & 0.005025 \\
L.DKMGP & $\mathbf{0.0096}$ & $\mathbf{0.001409}$ \\
VGP     & 0.0775 & 0.024445 \\
MVGP    & $\mathbf{0.0080}$ & $\mathbf{0.000643}$ \\
DGP     & 0.0615 & 0.004502 \\
DNN     & $\mathbf{0.0122}$ & $\mathbf{0.003068}$ \\
\bottomrule
\end{tabular}
\end{table}

\section{Posterior contraction: variance ratios}
\label{sec:supp-variance-ratios}

Table~\ref{tab:supp-variance-ratios} gives the parameter-specific posterior-to-prior variance ratios cited in the main manuscript. A smaller ratio indicates a greater reduction in marginal variance relative to the prior.

\begin{table}[htbp]
\centering
\caption{Mean posterior-to-prior variance ratio for the three shortlisted emulators. Zone indices 1-5 denote the anterior, inferior, lateral, septal and apical zones, respectively. Lower values indicate greater posterior contraction.}
\label{tab:supp-variance-ratios}
\begin{tabular}{@{}lccc@{}}
\toprule
Parameter & MVGP & L.DKMGP & DNN \\
\midrule
$C_a^{(1)}$ & 0.0073 & 0.7778 & 0.0039 \\
$C_b^{(1)}$ & 0.0139 & 0.9130 & 0.0078 \\
$C_a^{(2)}$ & 0.0143 & 0.7774 & 0.0071 \\
$C_b^{(2)}$ & 0.0351 & 0.9268 & 0.0152 \\
$C_a^{(3)}$ & 0.0046 & 0.6686 & 0.0022 \\
$C_b^{(3)}$ & 0.0122 & 0.8915 & 0.0061 \\
$C_a^{(4)}$ & 0.0076 & 0.6439 & 0.0029 \\
$C_b^{(4)}$ & 0.0285 & 0.9099 & 0.0114 \\
$C_a^{(5)}$ & 0.0031 & 0.6200 & 0.0017 \\
$C_b^{(5)}$ & 0.0096 & 0.8769 & 0.0056 \\
\bottomrule
\end{tabular}
\end{table}

\section{MCMC convergence diagnostics}
\label{sec:supp-convergence}

Tables~\ref{tab:supp-mcmc-diagnostics-mvgp}-
\ref{tab:supp-mcmc-diagnostics-dnn} provide the per-parameter convergence summaries underlying the aggregate statement in the main manuscript. For each model and parameter, the tables report the mean and median split $\widehat R$ and effective sample size (ESS) over the evaluated posterior fits. Values of $\widehat R$ close to one support between-chain consistency, while larger ESS values indicate more efficient posterior exploration.

\begin{longtable}{@{}llrrrr@{}}
\caption{Per-parameter MCMC diagnostic summaries for MVGP. Posterior fits used two chains with 600 retained draws and 300 warm-up iterations per chain.}
\label{tab:supp-mcmc-diagnostics-mvgp}\\
\toprule
Model & Parameter & \multicolumn{2}{c}{$\widehat R$} & \multicolumn{2}{c}{ESS} \\
\cmidrule(lr){3-4}\cmidrule(lr){5-6}
& & Mean & Median & Mean & Median \\
\midrule
\endfirsthead
\multicolumn{6}{c}{\tablename\ \thetable\ (continued)}\\
\toprule
Model & Parameter & \multicolumn{2}{c}{$\widehat R$} & \multicolumn{2}{c}{ESS} \\
\cmidrule(lr){3-4}\cmidrule(lr){5-6}
& & Mean & Median & Mean & Median \\
\midrule
\endhead
\midrule
\multicolumn{6}{r}{Continued on next page}\\
\endfoot
\bottomrule
\endlastfoot
MVGP & $C_a^{(1)}$ & 1.0008 & 0.9999 & 659.9 & 639.7 \\
MVGP & $C_b^{(1)}$ & 1.0007 & 0.9998 & 691.8 & 677.6 \\
MVGP & $C_a^{(2)}$ & 1.0011 & 1.0001 & 591.5 & 583.1 \\
MVGP & $C_b^{(2)}$ & 1.0009 & 0.9999 & 593.8 & 575.0 \\
MVGP & $C_a^{(3)}$ & 1.0003 & 0.9998 & 639.3 & 639.6 \\
MVGP & $C_b^{(3)}$ & 1.0003 & 0.9997 & 659.8 & 639.0 \\
MVGP & $C_a^{(4)}$ & 1.0143 & 0.9999 & 616.6 & 606.4 \\
MVGP & $C_b^{(4)}$ & 1.0009 & 0.9999 & 634.8 & 617.4 \\
MVGP & $C_a^{(5)}$ & 1.0007 & 1.0002 & 734.0 & 714.6 \\
MVGP & $C_b^{(5)}$ & 1.0009 & 0.9999 & 730.2 & 720.5 \\
\end{longtable}

\begin{longtable}{@{}llrrrr@{}}
\caption{Per-parameter MCMC diagnostic summaries for L.DKMGP. Posterior fits used two chains with 750 retained draws and 300 warm-up iterations per chain.}
\label{tab:supp-mcmc-diagnostics-ldkmgp}\\
\toprule
Model & Parameter & \multicolumn{2}{c}{$\widehat R$} & \multicolumn{2}{c}{ESS} \\
\cmidrule(lr){3-4}\cmidrule(lr){5-6}
& & Mean & Median & Mean & Median \\
\midrule
\endfirsthead
\multicolumn{6}{c}{\tablename\ \thetable\ (continued)}\\
\toprule
Model & Parameter & \multicolumn{2}{c}{$\widehat R$} & \multicolumn{2}{c}{ESS} \\
\cmidrule(lr){3-4}\cmidrule(lr){5-6}
& & Mean & Median & Mean & Median \\
\midrule
\endhead
\midrule
\multicolumn{6}{r}{Continued on next page}\\
\endfoot
\bottomrule
\endlastfoot
L.DKMGP & $C_a^{(1)}$ & 0.9999 & 0.9995 & 1957.0 & 1963.8 \\
L.DKMGP & $C_b^{(1)}$ & 0.9998 & 0.9996 & 2034.2 & 1965.5 \\
L.DKMGP & $C_a^{(2)}$ & 0.9999 & 0.9996 & 1941.8 & 1855.7 \\
L.DKMGP & $C_b^{(2)}$ & 0.9998 & 0.9995 & 1976.8 & 1875.4 \\
L.DKMGP & $C_a^{(3)}$ & 1.0000 & 0.9996 & 1833.5 & 1818.5 \\
L.DKMGP & $C_b^{(3)}$ & 1.0000 & 0.9996 & 1905.1 & 1848.3 \\
L.DKMGP & $C_a^{(4)}$ & 1.0000 & 0.9997 & 1823.3 & 1793.7 \\
L.DKMGP & $C_b^{(4)}$ & 0.9999 & 0.9995 & 1900.4 & 1899.8 \\
L.DKMGP & $C_a^{(5)}$ & 1.0000 & 0.9997 & 1789.6 & 1749.9 \\
L.DKMGP & $C_b^{(5)}$ & 0.9998 & 0.9995 & 1847.7 & 1788.8 \\
\end{longtable}

\begin{longtable}{@{}llrrrr@{}}
\caption{Per-parameter MCMC diagnostic summaries for DNN. Posterior fits used two chains with 600 retained draws and 300 warm-up iterations per chain.}
\label{tab:supp-mcmc-diagnostics-dnn}\\
\toprule
Model & Parameter & \multicolumn{2}{c}{$\widehat R$} & \multicolumn{2}{c}{ESS} \\
\cmidrule(lr){3-4}\cmidrule(lr){5-6}
& & Mean & Median & Mean & Median \\
\midrule
\endfirsthead
\multicolumn{6}{c}{\tablename\ \thetable\ (continued)}\\
\toprule
Model & Parameter & \multicolumn{2}{c}{$\widehat R$} & \multicolumn{2}{c}{ESS} \\
\cmidrule(lr){3-4}\cmidrule(lr){5-6}
& & Mean & Median & Mean & Median \\
\midrule
\endhead
\midrule
\multicolumn{6}{r}{Continued on next page}\\
\endfoot
\bottomrule
\endlastfoot
DNN & $C_a^{(1)}$ & 1.0010 & 1.0000 & 575.1 & 563.1 \\
DNN & $C_b^{(1)}$ & 1.0009 & 0.9998 & 624.6 & 612.8 \\
DNN & $C_a^{(2)}$ & 1.0019 & 1.0003 & 481.4 & 474.7 \\
DNN & $C_b^{(2)}$ & 1.0018 & 1.0005 & 490.6 & 484.7 \\
DNN & $C_a^{(3)}$ & 1.0015 & 1.0002 & 532.5 & 521.7 \\
DNN & $C_b^{(3)}$ & 1.0015 & 1.0001 & 558.5 & 545.8 \\
DNN & $C_a^{(4)}$ & 1.0008 & 1.0001 & 518.0 & 515.9 \\
DNN & $C_b^{(4)}$ & 1.0008 & 0.9999 & 533.0 & 530.5 \\
DNN & $C_a^{(5)}$ & 1.0010 & 0.9998 & 603.0 & 603.3 \\
DNN & $C_b^{(5)}$ & 1.0011 & 0.9999 & 592.8 & 584.3 \\
\end{longtable}

Across all model-parameter combinations, the median $\widehat R$ values range from 0.9995 to 1.0005, and all mean ESS values exceed 400. The mean $\widehat R$ for $C_a^{(4)}$ under MVGP is 1.0143 despite a median of 0.9999. This difference indicates that the median alone does not describe the full distribution of the diagnostic across the evaluated fits.

\section{Complete posterior trajectories and marginal distributions}
\label{sec:supp-full-posteriors}

\begin{figure}[!t]%
  \centering
  \includegraphics[width=1\textwidth, keepaspectratio]{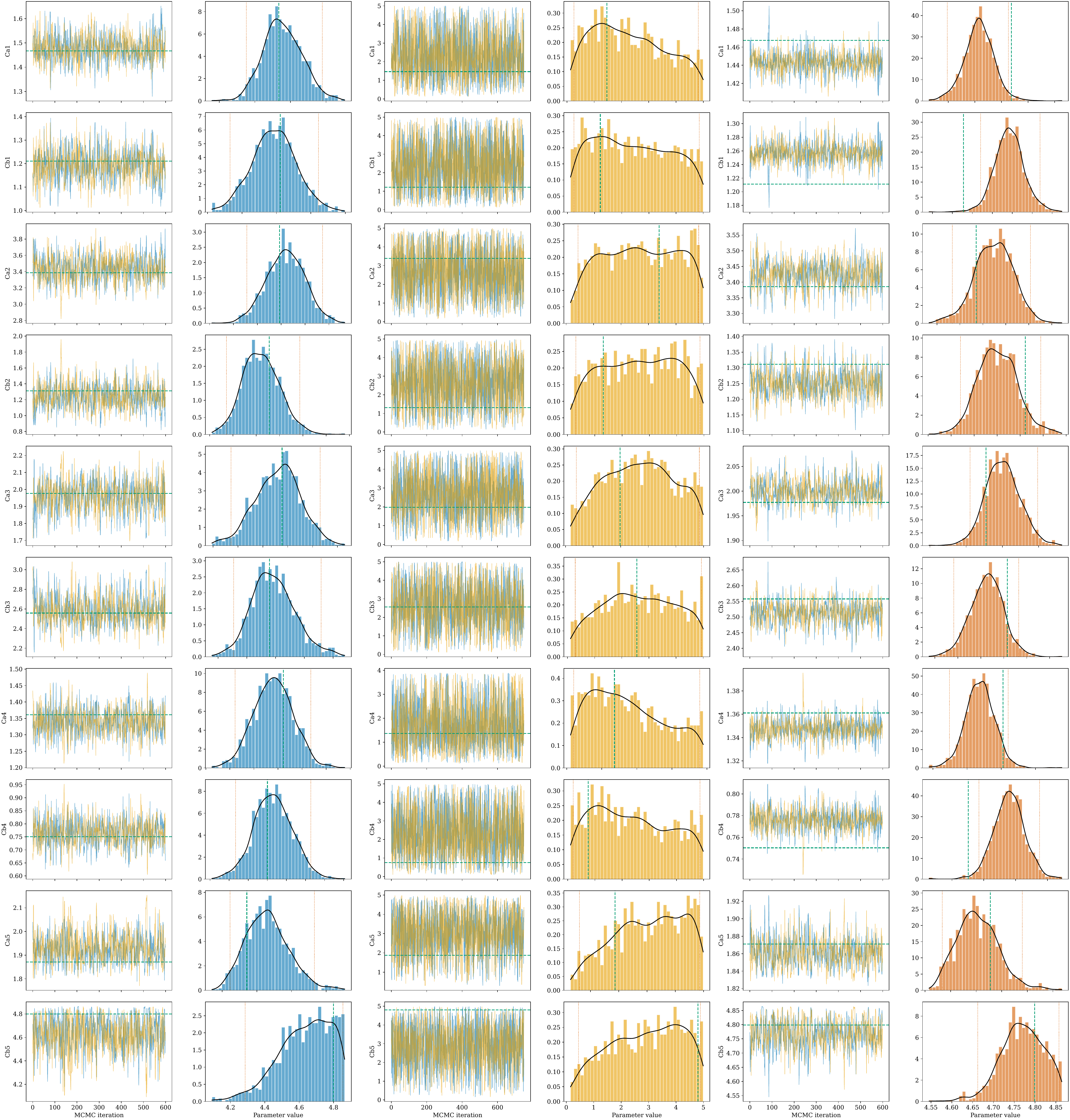}
  \caption{MCMC trace plots and posterior distributions for a representative test sample across the three top-ranked emulators: MVGP (left), L.DKMGP (centre), and DNN (right). Each row corresponds to one of the ten Holzapfel-Ogden parameters ($Ca_1$-$Cb_5$). Dashed horizontal/vertical lines indicate the true parameter values used in data generation, and results are paired trace and posterior-density panels. Trace plots display two MCMC chains and the true parameter value. Posterior panels show samples, kernel density estimates, the true value, and 95\% credible intervals.}
  \label{fig:supp-representative-posteriors}
\end{figure}

Figure~\ref{fig:supp-representative-posteriors} provides the complete chain trajectories and marginal posterior distributions for the representative held-out test sample cited in the main manuscript. These detailed diagnostics are not reproduced in the main manuscript.

\section{Full posterior parameter-correlation matrices}
\label{sec:supp-intra-parameter-correlation}

\begin{figure}[p]
\centering
\includegraphics[width=0.9\textwidth, keepaspectratio]{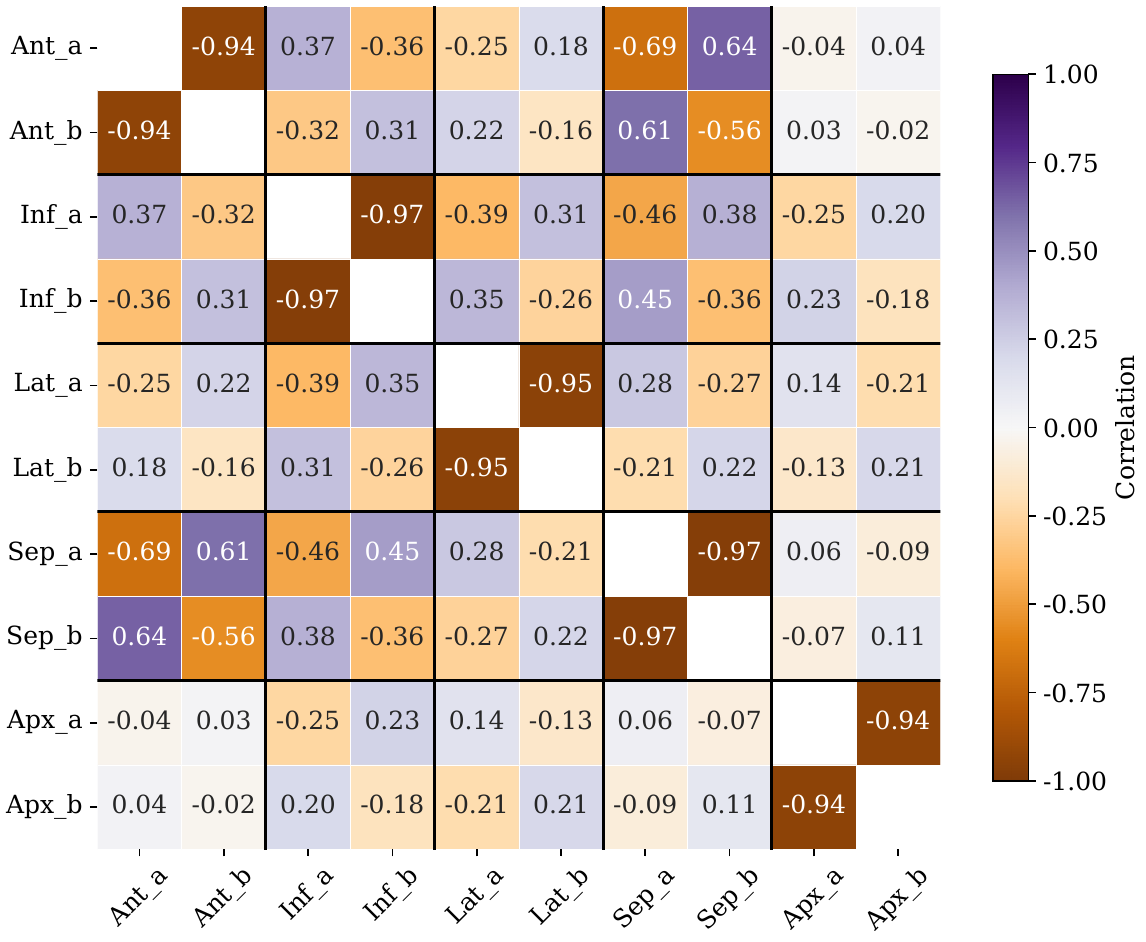}
\caption{%
Full posterior parameter-correlation matrix for the multi-territory stiffening scenario $(\mu=1.7)$. All five anatomical zones are stiffened simultaneously. Each off-diagonal cell reports the mean posterior correlation between the indicated pair of regional parameters over the 100 test samples in this scenario. The colour scale ranges from $-1$, indicating perfect negative correlation, to $1$, indicating perfect positive correlation.
}
\label{fig:supp-correlation-multiterritory}

\end{figure}

\begin{figure}[p]
\centering
\includegraphics[width=0.9\textwidth, keepaspectratio]{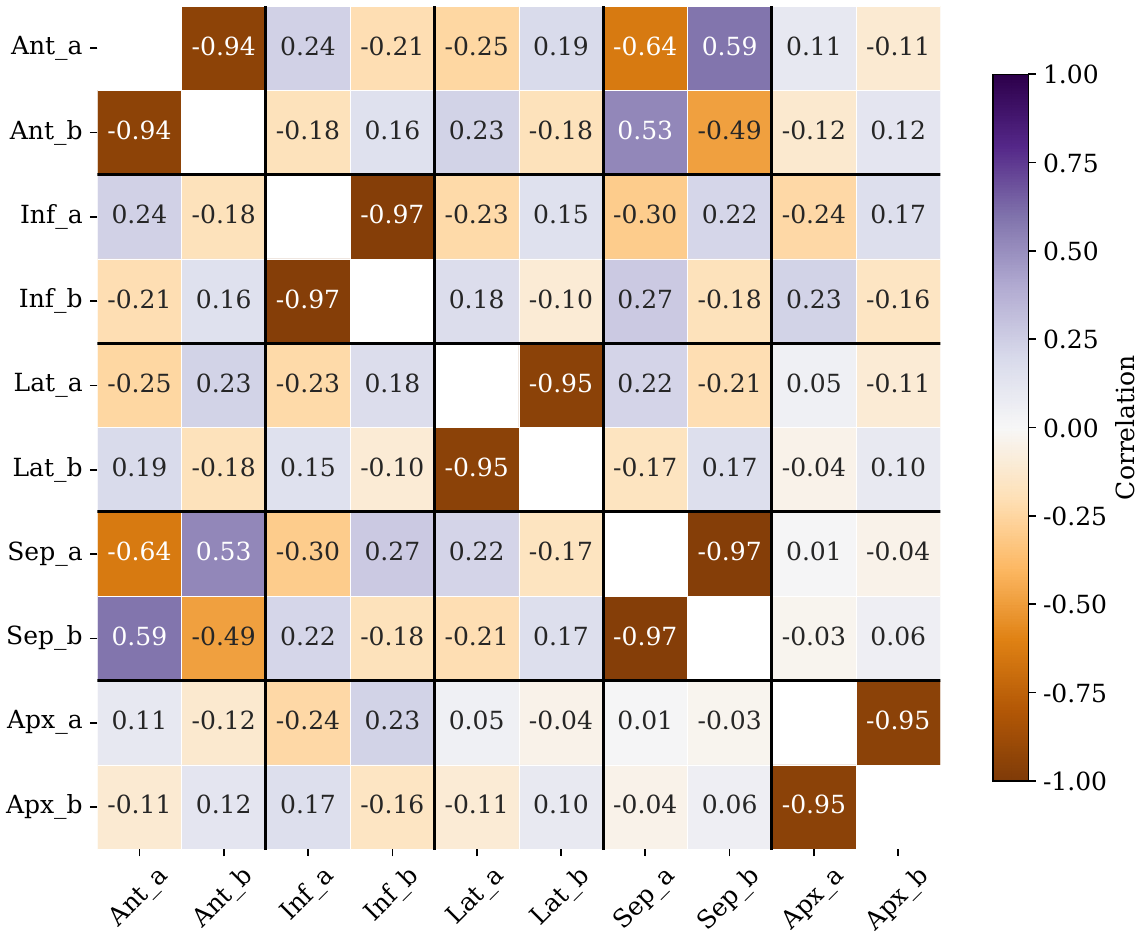}%
\caption{%
Full posterior parameter-correlation matrix for the mild local-stiffening scenario $(\mu=1.5)$. One anatomical zone is stiffened in each test case. Each off-diagonal cell reports the mean posterior correlation between the indicated pair of regional parameters over the 100 test samples in this scenario. The colour scale ranges from $-1$, indicating perfect negative correlation, to $1$, indicating perfect positive correlation.
}
\label{fig:supp-correlation-mild}

\end{figure}

\begin{figure}[p]
\centering
\includegraphics[width=0.9\textwidth, keepaspectratio]{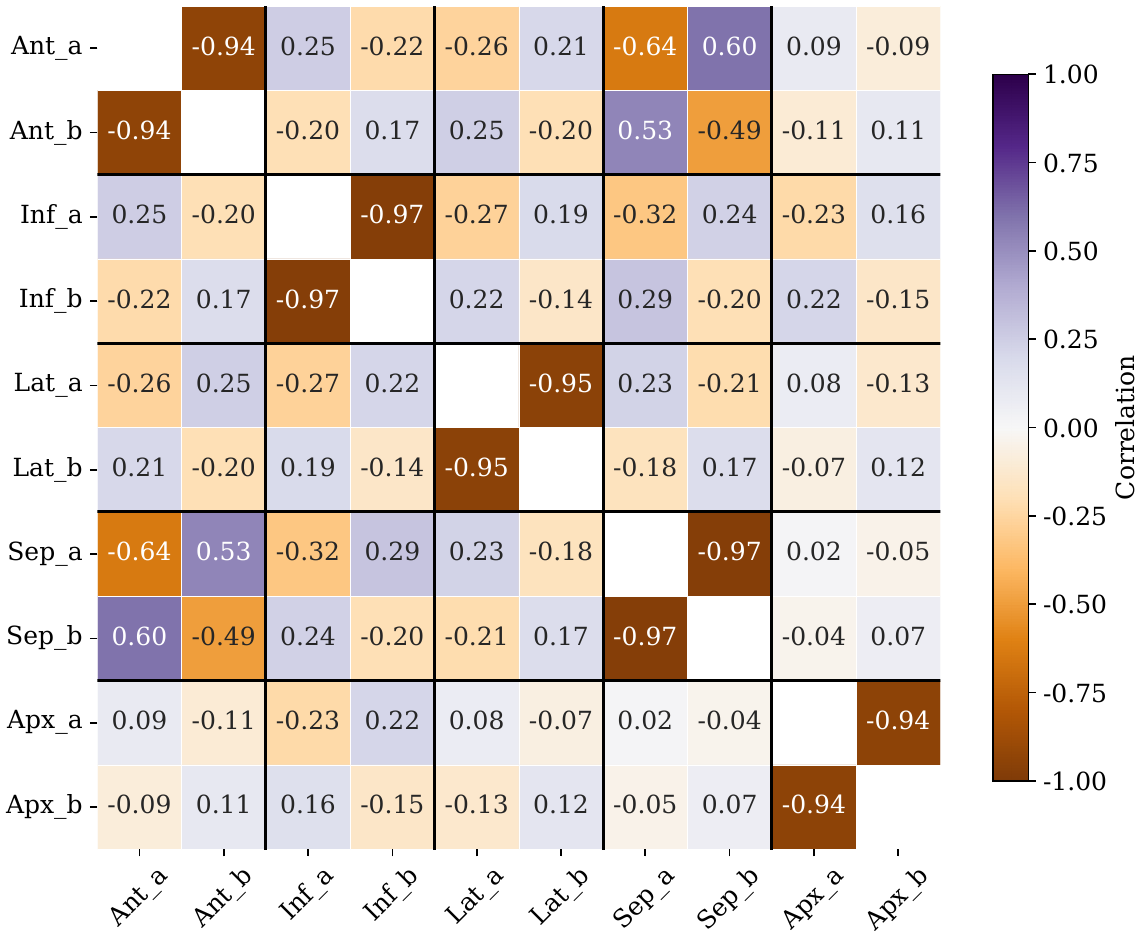}%

\caption{%
Full posterior parameter-correlation matrix for the moderate local-stiffening scenario $(\mu=2.0)$. One anatomical zone is stiffened in each test case. Each off-diagonal cell reports the mean posterior correlation between the indicated pair of regional parameters over the 100 test samples in this scenario. The colour scale ranges from $-1$, indicating perfect negative correlation, to $1$, indicating perfect positive correlation.
}
\label{fig:supp-correlation-moderate}

\end{figure}

\begin{figure}[p]
\centering
\includegraphics[width=0.9\textwidth, keepaspectratio]{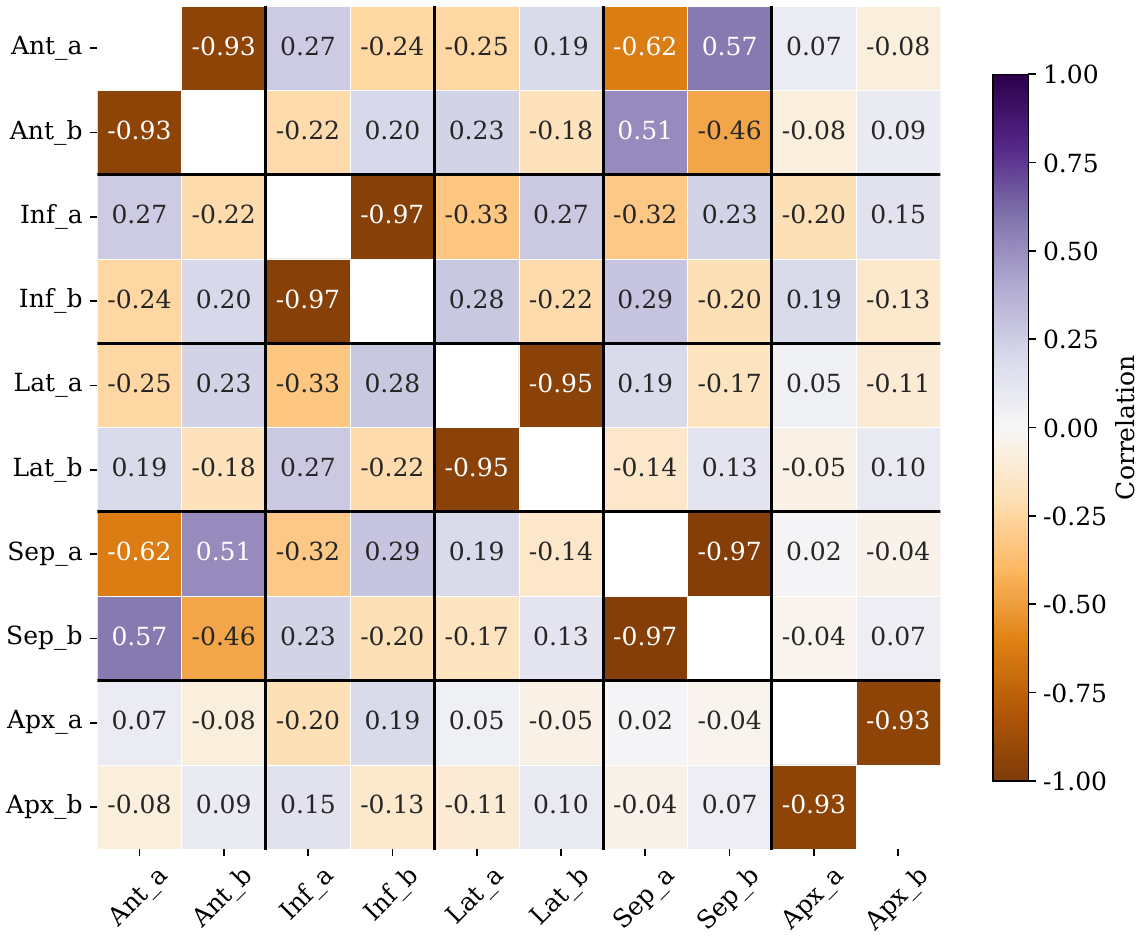}%
\caption{%
Full posterior parameter-correlation matrix for the severe local-stiffening scenario $(\mu=3.0)$. One anatomical zone is stiffened in each test case. Each off-diagonal cell reports the mean posterior correlation between the indicated pair of regional parameters over the 100 test samples in this scenario. The colour scale ranges from $-1$, indicating perfect negative correlation, to $1$, indicating perfect positive correlation.
}
\label{fig:supp-correlation-severe}

\end{figure}

Figures~\ref{fig:supp-correlation-multiterritory}-\ref{fig:supp-correlation-severe} present the full $10\times10$ posterior parameter-correlation matrices before the within-zone first-principal-component projection used for the inter-regional summaries in the main manuscript. The matrices therefore retain both the dependence between $C_a^{(z)}$ and $C_b^{(z)}$ within each zone and the dependence between parameters belonging to different zones. The displayed correlations were averaged over the 100 test samples in each synthetic stiffening scenario.

The row and column abbreviations Ant, Inf, Lat, Sep and Apx denote the anterior, inferior, lateral, septal and apical zones, respectively. The suffixes \texttt{\_a} and \texttt{\_b} identify the corresponding $C_a$- and $C_b$-type parameters. Negative correlations are shown in orange-brown, positive correlations in purple, and correlations close to zero in pale colours. The main diagonal is omitted.

\section{Clinical application: MCMC diagnostics}
\label{sec:supp-clinical-diagnostics}

\begin{figure}[p]
\centering
\includegraphics[width=1\textwidth, keepaspectratio]{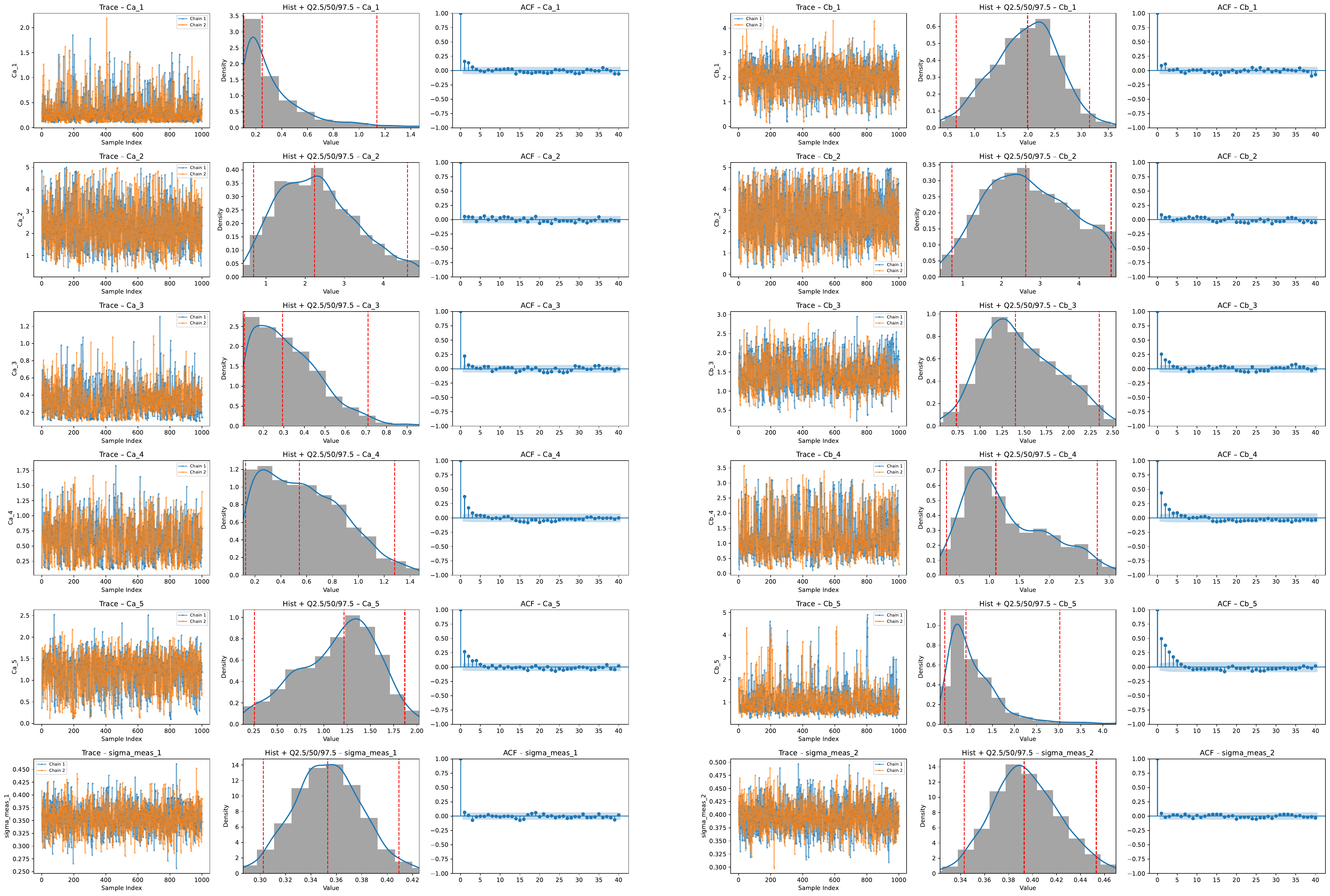}%
\caption{
Complete MCMC diagnostics for the clinical analysis. Each row corresponds to one of the ten regional constitutive parameters or one of the two measurement-noise scales. The left column shows the retained draws from the two chains, with chain~1 in blue and chain~2 in orange. The centre column shows the marginal posterior histogram in grey and the kernel density estimate in blue; the red dashed lines mark the 2.5th, 50th and 97.5th posterior percentiles. The right column shows the autocorrelation function through lag~40, with the shaded region providing the reference band around zero. The parameters denoted by $\sigma_1$ and $\sigma_2$ are the circumferential- and radial-strain measurement-noise scales, respectively.
}
\label{fig:supp-clinical-mcmc}
\end{figure}

Figure~\ref{fig:supp-clinical-mcmc} provides the full sampling diagnostics cited in the clinical application. The posterior summaries and posterior predictive results reported in the main manuscript are not repeated here.

Table~\ref{tab:supp-clinical-convergence} supplements the graphical diagnostics in Figure~\ref{fig:supp-clinical-mcmc} with parameter-specific convergence summaries. Across the twelve inferred parameters, the reported $\widehat R$ values range from $1.000$ to $1.002$, the split-$\widehat R$ values range from $0.999$ to $1.004$, and the effective sample sizes range from $441.5$ to $1762.9$. 

\begin{table}[htbp]
\centering
\caption{Parameter-specific MCMC convergence summaries for the clinical analysis. The columns report the conventional potential scale reduction factor $\widehat R$, its split-chain version, and the reported ESS.
}
\label{tab:supp-clinical-convergence}
\begin{tabular}{@{}lrrr@{}}
\toprule
Parameter
  & $\widehat R$
  & Split-$\widehat R$
  & ESS \\
\midrule
$C_a^{(1)}$ & 1.000 & 1.000 & 1061.8 \\
$C_b^{(1)}$ & 1.000 & 0.999 & 1381.7 \\
$C_a^{(2)}$ & 1.000 & 1.001 & 1646.9 \\
$C_b^{(2)}$ & 1.000 & 0.999 & 1538.0 \\
$C_a^{(3)}$ & 1.000 & 1.000 & 1021.0 \\
$C_b^{(3)}$ & 1.002 & 1.004 &  811.3 \\
$C_a^{(4)}$ & 1.000 & 1.002 &  859.9 \\
$C_b^{(4)}$ & 1.000 & 1.001 &  711.3 \\
$C_a^{(5)}$ & 1.000 & 1.001 &  728.5 \\
$C_b^{(5)}$ & 1.000 & 1.003 &  441.5 \\
$\sigma_1$  & 1.000 & 0.999 & 1762.9 \\
$\sigma_2$  & 1.000 & 1.001 & 1563.9 \\
\bottomrule
\end{tabular}
\end{table}

\end{document}